\documentclass[preprint,showpacs,preprintnumbers,amsmath,amssymb]{revtex4}
\usepackage{graphicx}
\usepackage{dcolumn}
\usepackage{bm}
\usepackage[]{amsmath,amssymb,epsfig,color}
\newcommand{\be}{\begin{equation}}
\newcommand{\ee}{\end{equation}}
\begin{document}
\title{Stochastic dynamics and control  of a driven nonlinear spin chain: the role of  Arnold diffusion}

\author{L. Chotorlishvili$^{1,2,3}$, Z.Toklikishvili$^{3}$, J. Berakdar$^{1}$}
%
%
\affiliation{1 Institut f\"ur Physik, Martin-Luther Universit\"at
Halle-Wittenberg, Heinrich-Damerow-Str.4, 06120 Halle, Germany \\
  2 Institut f\"ur Physik, Universit\"at Augsburg, 86135 Augsburg,Germany\\
  3 Physics Department of the Tbilisi State University,
                 Chavchavadze av.3, 0128, Tbilisi}

\date{\today}
\begin{abstract}
We study a chain of non-linear, interacting spins driven by a static and
a time-dependent magnetic field. The aim is to identify the conditions
for the locally and temporally controlled spin switching.
Analytical and  full numerical calculations show the possibility
of stochastic control if the underlying semi-classical dynamics is
chaotic. This is achievable by tuning the external field parameters according
to the method described in this paper.
We show analytically for a finite spin chain that Arnold diffusion is the underlying
mechanism for the present stochastic control.
Quantum mechanically we consider the regime where the
classical dynamics is regular or chaotic. For the latter we utilize
 the random matrix theory. The efficiency and  the stability of the
non-equilibrium quantum spin-states are quantified by the
time-dependence of the Bargmann angle related to the geometric phases of the states.
\end{abstract}
\maketitle

\section{Introduction }
Advances in nanoscale fabrication of magnetic materials down to
a finite chain of individual magnetic atoms \cite{science} triggered
a number of studies on the ground state magnetic properties of finite, interacting
spin chains \cite{spintheory}.
For accessing the non-equilibrium states,   in a linear chain one  conventionally rotates
the spins  by
 applying a static magnetic field $\mathbf H_{0}$ and a
variable magnetic field $\mathbf h(t)$ along a   direction
perpendicular to  $H_{0}$  \cite{Abragam}. The spins affected by
the fields are then deflected by an angle  $\theta=\omega_{1}\tau$
which can be desirably varied by changing the duration  $\tau$ of
the field $\mathbf h(t)$. Here $\omega_{1}$ is the amplitude of
$\mathbf h(t)$  in frequency units $\omega_{1}=h_{0}\gamma$
($\gamma$ is the gyromagnetic ratio).  For this scheme to be viable
 $\omega_{1}$ has to be in resonance with the system  precessional
frequency. In this paper we consider the spin deflection in the
different situation of a  nonlinear chain of interacting spins
\cite{Mejia-Monasterio,Saito} in which case the precessional
frequency is dynamical and changes with the oscillation
 amplitude \cite{Sagdeev}. Hence, a control strategy \cite{Matos-Abiague,Matos-Abiague2}
 as in the linear chain case entails the
 use of chirped fields. Here we inspect a different route to spin control
  by exploiting the stochastic nature
 of the spin dynamics when appropriate fields are employed. This we show in a first step analytically. The advantage is that no special
 frequency tuning is used and more importantly the spin may be quasi stable at the
  deflected (non-equilibrium) angles when the field $\mathbf h(t)$  is off which might be of interest for
  quantum information applications \cite{Yuan,Tuchette,Mabuchi,Hood,Raimond,Chotorlishvili,Lakshmanan,Zolotaryuk}.
  Disadvantage is the limited control of the switching time.
Full numerical simulations confirm our analytical predictions: Tuning the  external
 fields such that the underlying classical spin dynamics is chaotic,
   stochastic switching occurs  and
  a long-time  quasi stabilization, i.e. a dynamical freezing (DF) of the deflected states is possible.
Small fields cause only small fluctuations
around the equilibrium state.
For very strong fields, effects of magnetic anisotropy and exchange become subsidiary and hence
the dynamics turn regular and no deflection with subsequent freezing is possible.

For a finite spin chain we uncover analytically that our stochastic control (SC) scheme
is governed by Arnold diffusion and give analytical expression for the Arnold diffusion coefficient that in turn
determine the time scale for SC.

To inspect the influence of the quantum nature of the spins on our (classical) predictions
we considered both the regular and the chaotic classical regimes and evaluated the so-called
Bargmann angle which is a measure of the quantum distance between states in the Hilbert space
and can be used to signal DF \cite{Matos-Abiague,Matos-Abiague2}.
 Using random matrix theory we prove indeed that SC and DF are possible at the driving field values
 that follows from our classical analysis.
\begin{figure}[t]
 \centering
  \includegraphics[width=12cm]{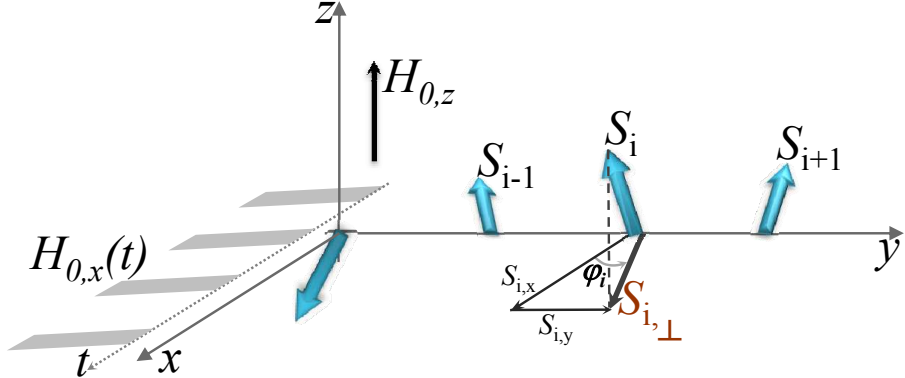}
  \caption{A schematics  of the interacting spin chain. The magnetic anisotropy field
  sets the $z$ direciton. Two magnetic fields are applied: a static ($H_{0,z}$) one along the
  $z$ axis  and
   a time-dependent field ($H_{0,x}(t)$) along the
  $x$ axis. } \label{Fig:1}
\end{figure}
\section{Equation of motion}

\subsection{Liouville equation for the spin chain}

 Similar to the case studied
in \cite{Mejia-Monasterio,Saito} we consider a system that can be
modeled by  a chain of $N$  interacting (with coupling constant
$J$)
 spin variables
localized at sites $j$ and having a uniaxial anisotropy with the anisotropy constant $\beta$. Possible sources of the anisotropy field  are discussed in Refs.\cite{science}. Here we mention that the inclusion of a finite
anisotropy is essential for the existence of  a finite-temperature long-range order in the (infinitely long) chain. Further important
consequences of the magnetic anisotropy for the phenomena discussed in this work are detailed below.
The direction
of the anisotropy
field defines the $z$ axis.
A static ($H_{0z}$) and  a time-dependent ($H_{0x}(t)$) magnetic fields
  are applied along the $z$ and $x$ axes, respectively (cf. Fig.\ref{Fig:1}).
 $H_{0x}(t)$ consists of $N_k$ periodic pulses, i.e.
\begin{equation}
H_{0x}(t)=\varepsilon
T\sum\limits_{k}^{N_k}\delta(t-kT),
\label{eq:field}\end{equation}
where $T$ is the period and $\varepsilon T$  is the field strength. As demonstrated
explicitly \cite{prl09} (for the classical spin dynamics)
the shape (\ref{eq:field}) of the field  mimics well the action of a
 finite-width pulse as long as the pulse duration  is smaller than the
 field-free precessional period of the spins. The time-integral over the
 field amplitude of the finite-width pulse sets the variables $\varepsilon$ \cite{prl09}.

From the  Hamilton operator
\cite{Mejia-Monasterio,Saito}
\begin{equation}
\hat{H}=\sum\limits_{j=1}^{N-1}J \hat s_{jz} \hat
s_{j+1,z}+H_{0x}\sum\limits_{j=1}^{N}\hat
s_{jx}+H_{0z}\sum\limits_{j=1}^{N}\hat s_{jz}+\beta
\sum\limits_{j=1}^{N}\hat s^{2}_{jz} \label{eq:ham}\end{equation}
we find the spin equation of motion (EOM) to be
\begin{equation}
\partial_t \mathbf{\hat s}_{j} =  \mathbf{h}_j\times \mathbf{\hat s}_{j},\quad
\mathbf{ h}_{j}=\left(H_{0x},0,J[{\hat s}_{j+1 z}+ {\hat s}_{j-1
z}]+H_{0z}+2\beta \hat s_{jz} \right). \label{eq:h}
\end{equation}
For large spins $s_{j}$
 with $\hbar s_{j}(s_j+1)=\langle  \mathbf{{\hat s}}_{j} \rangle$ and
 $[\hat{H},\mathbf{\hat s}_{i}^2]=0$
 we shift variables as (cf. Fig.\ref{Fig:1}, $s_i^2=1$)
\begin{equation}
 s_{ix}=s_{i\perp}\cos\varphi_{i};~~~s_{iy}=s_{i\perp}\sin\varphi_{i},
 \ s_{i,\perp}=\sqrt{1-s_{i,z}^{2}} .
 \label{eq:para}
 \end{equation}
EOM for two spins $N=2$  in term of the canonical
action variables   $s_{1z},$ $s_{2z}$ and their conjugate angles
$\varphi_{1},$ $\varphi_{2}$ reads
\begin{eqnarray}\label{eq:conv}
\dot{s}_{iz}&=&-\varepsilon\frac{\partial V}{\partial
\varphi_{i}},\quad  \dot{\varphi}_{i}=\omega_{i}(s_{jz})+\varepsilon\frac{\partial
V}{\partial s_{iz}},\quad \omega_{i}(s_{jz})=Js_{jz}+2\beta \hat s_{iz}+H_{0z},\quad i,j\in\{1,2\};\nonumber\\
V&:=&\sum\limits_{i=1}^{N}V_{i}(s_{iz},\varphi_{i})=\sum\limits_{i=1}^{N}V_{0i}(s_{iz},\varphi_{i})\
T\sum\limits_{k=-\infty}^{+\infty}\delta(t-kT)
=\sum\limits_{i=1}^{N}s_{i\perp}\cos \varphi_{i}\cdot T\sum\limits_{k=-\infty}^{+\infty}\delta(t-kT).\nonumber\\
\end{eqnarray}
For $V\neq0$ the variables of actions are adiabatic invariants and
hence are slow with respect to the angles typical time scale
\cite{Sagdeev}. The idea now is to identify the regime of
classical chaotic dynamics which we will do below. In this regime
one may adopt a kinetic approach based on the  Liouville  equation
\cite{Balescu,Lichtenberg}  for the two-particle distribution
function $f(t,s_{1z},\varphi_{1},s_{2z},\varphi_{2})$, i.e.\
\begin{eqnarray}
 i\frac{\partial f}{\partial t}&=&\big(\hat{L}_{0}+\varepsilon
\hat{L}_{1}\big)f,\nonumber\\
\hat{L}_{0}&=&-i\omega_{1}(s_{2z})\frac{\partial }{\partial
\varphi_{1}}-i\omega_{2}(s_{1z})\frac{\partial }{\partial
\varphi_{2}},\nonumber\\
\hat{L}_{1}&=&-i\Bigg(\frac{\partial V}{\partial
s_{1z}}\, \frac{\partial }{\partial \varphi_{1}}-\frac{\partial
V}{\partial \varphi_{1}}\, \frac{\partial }{\partial
s_{1z}}\Bigg)-i\Bigg(\frac{\partial V}{\partial
s_{2z}}\, \frac{\partial }{\partial \varphi_{2}}-\frac{\partial
V}{\partial \varphi_{2}}\, \frac{\partial }{\partial
s_{2z}}\Bigg).
\end{eqnarray}
Eq.(6) is of a key importance for this study. Below we
use the random phase approximation and some mathematical
techniques  to derive from eg.(6) the Fokker-Planck equation which allows
 to explore some chaotic features of the spin dynamics.

\subsection{Fokker-Planck formulation and the onset of the chaotic regime}

Expressing  $f(s_{1z},\varphi_{1},s_{2z},\varphi_{2})$ as a
Fourier series over $\varphi_{1}$ and $\varphi_{2}$ we find
\begin{eqnarray}
f(s_{1z},\varphi_{1},s_{2z},\varphi_{2})&=&
\frac{1}{(2\pi)^{2}}\sum_{m,n}\bar{f}_{n,m}(s_{1z},s_{2z})
e^{in\varphi_{1}}e^{im\varphi_{2}},\nonumber\\
\bar{f}_{n,m}(s_{1z},s_{2z})&=&f_{n,m}(s_{1z},s_{2z})\cdot
\exp\big[-in\int\limits_{0}^{t}\omega_{1}(t')dt'\big]
\cdot\exp\big[-im\int\limits_{0}^{t}\omega_{2}(t')dt'\big],\nonumber\\
\omega_{1}\big(t'\big)&=&\omega_{1}\big(s_{2z}(t')\big),\omega_{2}\big(t'\big)=\omega_{2}\big(s_{1z}(t')\big).
\end{eqnarray}
Hence,  solution of the
Liouville equation is cast formally as (the symbol $\langle n',m'|\hat{L}_{1}(t_{1})|n,m\rangle$ means the average
over fast oscillating variables)
\begin{eqnarray}
&&f_{n',m'}\big(s_{1z},s_{2z},t\big)=f_{n',m'}\big(s_{1z},s_{2z},0\big)-
\nonumber \\
&&-i\varepsilon\sum_{n,m}\int\limits_{0}^{t}dt_{1}e^{i(n'-n)\int\limits_{0}^{t_{1}}\omega_{1}(t')dt'}e^{i(m'-m)\int\limits_{0}^{t_{1}}\omega_{2}(t')dt'}\langle
n'm'|\hat{L}_{1}(t_{1})|n,m\rangle
f_{n,m}\big(s_{1z},s_{2z},t_{1}\big).
\end{eqnarray}
If the interaction energy with the variable field is small with
respect to the other terms in eq.(\ref{eq:ham}) we can expand $f$
in terms of the field strength $\varepsilon$ and account for
leading terms only.

The zero-order component has the form

\begin{eqnarray}\label{eq:f00}
&&f_{0,0}\big(s_{1z},s_{2z},t\big)=f_{0,0}\big(s_{1z},s_{2z},0\big)-
\nonumber \\
&&-i\varepsilon\sum_{n,m}\int\limits_{0}^{t}dt_{1}e^{-in
\int\limits_{0}^{t_{1}}\omega_{1}(t')dt'}e^{-im\int\limits_{0}^{t_{1}}\omega_{2}(t')dt'}\langle
0,0|\hat{L}_{1}(t_{1})|n,m\rangle
f_{n,m}\big(s_{1z},s_{2z},0\big)+ \\
&&+(-i\varepsilon)^{2}\sum_{n,m}\int\limits_{0}^{t}dt_{1}
\int\limits_{0}^{t_{1}}dt_{2}e^{in\int\limits_{t_{1}}^{t_{2}}
\omega_{1}(t')dt'}e^{im\int\limits_{t_{1}}^{t_{2}}\omega_{2}(t')dt'}
\langle 0,0|\hat{L}_{1}(t_{1})|n,m\rangle \langle
n,m|\hat{L}_{1}(t_{2})|0,0\rangle f_{0,0}\big(s_{1z},s_{2z},0\big).
\nonumber
\end{eqnarray}
Now we write $\hat{L}_{1}$ as a Fourier series taking the relevant
frequency
$\Omega=\frac{2\pi}{T}$  into account,
$\hat{L}_{1}(t)=\sum_{p}L_{1,p}\exp(ip\Omega t),
L_{1,-p}=L_{1,p}^{\ast} $. Inserting into (9) we find
\begin{eqnarray}
&&f_{0,0}\big(s_{1z},s_{2z},t\big)=f_{0,0}\big(s_{1z},s_{2z},0\big)-
\nonumber \\
&&-i\varepsilon\sum_{n,m}\int\limits_{0}^{t}dt_{1}e^{-in\int\limits_{0}^{t_{1}}\omega_{1}(t')dt'}e^{-im\int\limits_{0}^{t_{1}}\omega_{2}(t')dt'}\langle
0,0|\hat{L}_{1}(t_{1})|n,m\rangle
f_{n,m}\big(s_{1z},s_{2z},0\big)+ \\
&&+(-i\varepsilon)^{2}\sum_{n,m,p}\int\limits_{0}^{t}dt_{1}\int\limits_{0}^{t_{1}}dt_{2}e^{in\int\limits_{t_{1}}^{t_{2}}\omega_{1}(t')dt'}e^{im\int\limits_{t_{1}}^{t_{2}}\omega_{2}(t')dt'}
\langle 0,0|\hat{L}_{1,p}(t_{1})|n,m\rangle \langle
n,m|\hat{L}_{1,-p}(t_{2})|0,0\rangle\, \nonumber \\
&& e^{ip\Omega(t_{1}-t_{2})} f_{0,0}\big(s_{1z},s_{2z},0\big).
\nonumber
\end{eqnarray}
Introducing the  notations
\begin{equation}
\psi_{1}(t_{1},t_{2})=\int\limits_{t_{1}}^{t_{2}}\omega_{1}(t')dt',\;
\psi_{2}(t_{1},t_{2})=\int\limits_{t_{1}}^{t_{2}}\omega_{2}(t')dt'
\end{equation}
we write
\begin{eqnarray} \label{eq:20}
&&f_{0,0}\big(s_{1z},s_{2z},t\big)=f_{0,0}\big(s_{1z},s_{2z},0\big)-
\nonumber \\
&&-i\varepsilon\sum_{n,m}\int\limits_{0}^{t}dt_{1}e^{-in\psi_{1}(t_{1},0)}e^{-im\psi_{2}(t_{1},0)}\langle
0,0|\hat{L}_{1}(t_{1})|n,m\rangle
f_{n,m}\big(s_{1z},s_{2z},0\big)+ \\
&&+(-i\varepsilon)^{2}\sum_{n,m,p}\int\limits_{0}^{t}dt_{1}\int\limits_{0}^{t_{1}}dt_{2}e^{-in\psi_{1}(t_{1},t_{2})}e^{-im\psi_{2}(t_{1},t_{2})}
\langle 0,0|\hat{L}_{1,p}|n,m\rangle \langle
n,m|\hat{L}_{1,-p}|0,0\rangle \cdot\nonumber \\
&& \cdot e^{ip\Omega(t_{1}-t_{2})}
f_{0,0}\big(s_{1z},s_{2z},0\big). \nonumber
\end{eqnarray}
 Averaging over the initial
phases, i.e. $F\big(s_{1z},s_{2z},t\big)=\Bigg\langle\Big\langle
f_{0,0}\big(s_{1z},s_{2z},t\big)\Big\rangle\Bigg\rangle$, and
using the random phase approximation
$\psi_{1}(t_{1},t_{2})= \int\limits
_{t_{1}}^{t_{2}}\omega_{1}(t')dt'\approx\varphi_{1}(t_{2})-\varphi_{1}(t_{2}),$
we end up with  \begin{equation}
\Bigg\langle\Big\langle\exp\big[in\psi_{1,2}(t_{2},t_{1})\big]\Big\rangle\Bigg\rangle=
\exp\big(-(t_{1}-t_{2})/\tau_{c}\big)\exp\big(-in\omega_{1,2}(t_{1}-t_{2})\big).
\end{equation}
Here $\tau_{c}$ is the correlation time of random phase. Taking
eq.(12) into account we deduce then for the averaged two-particle
distribution function $F(t)$ the dynamical equation (up to a
second order in the field strength $\varepsilon$)
 \begin{eqnarray} \label{eq:part}
 &&\frac{\partial F}{\partial
t}=-i\varepsilon
e^{-\frac{2t}{\tau_{c}}}\sum\limits_{n,m}e^{-i(n\omega_{1}+m\omega_{2})t}\langle
0,0|\hat{L}_{1}(t)|n,m\rangle f_{n,m}(s_{1z},s_{2z},0)- \nonumber \\
&&-\varepsilon^{2}\frac{\partial}{\partial
t}\sum\limits_{m,n,p}\int\limits_{0}^{t}dt_{1}\int\limits_{0}^{t_{1}}dt_{2}e^{-\frac{2(t_{1}-t_{2})}
{\tau_{c}}}e^{-in\omega_{1}(t_{1}-t_{2})}e^{-im\omega_{2}(t_{1}-t_{2})}e^{ip\Omega(t_{1}-t_{2})}\\
&& \langle0,0|L_{1,p}|n,m\rangle\langle m,n|L_{1,-p}|0,0\rangle
F(s_{1z},s_{2z},t).\nonumber
\end{eqnarray}
%
The long time behaviour ($t\gg \tau_c$) is retrieved by shifting to the new variables
 $\tau=t_{1}-t_{2}$,$t_{1}=t_{1}$ in (14) and integrating over
  $\tau$  which yields
 \begin{equation} \label{eq:prt}
 \frac{\partial F}{\partial t}=-\varepsilon^{2}\sum\limits_{n,m,p}\frac{1}{\frac{2}{\tau_{c}}+i(n\omega_{1}+m\omega_{2}-p\Omega)}
 \langle 0,0 |L_{1p}|n,m\rangle\langle m,n|L_{1,-p}|0,0\rangle
 F(s_{1z},s_{2z},t).
 \end{equation}
For a further progress  explicit expressions for the matrix elements $\langle
0,0 |L_{1,p}|n,m\rangle\langle m,n|L_{1,-p}|0,0\rangle$ are need.
 Following the standard procedure outlined in  \cite{Haken}  we find after some lengthy steps
 the following Fokker-Planck equation  for  $F(s_{1z},s_{2z}, t)$
\begin{equation}\label{eq:planck}
\frac{\partial F(s_{1z},s_{2z},t)}{\partial t}=D\Bigg(
\frac{\partial}{\partial s_{1z}}(1-s_{1z}^{2})\frac{\partial
F}{\partial s_{1z}}+\frac{\partial}{\partial
s_{2z}}(1-s_{2z}^{2})\frac{\partial F}{\partial
s_{2z}}\Bigg),\quad
D=\frac{\varepsilon^{2}\pi}{2\Omega}=\frac{\varepsilon^{2} T}{4}.\end{equation}
Making  the ansatz
$ F(s_{1z},s_{2z})=F_{1}(s_{1z})F_{2}(s_{2z})$
 the average values of the spin projections
$ \bar{s}_{jz}$ is determined from
\begin{eqnarray}
\frac{d}{dt}\bar{s}_{jz}&=&\int\limits_{-1}^{+1}s_{jz}\frac{\partial
F_{j}}{\partial
t}ds_{jz}=D\int\limits_{-1}^{+1}s_{1z}\frac{\partial}{\partial
s_{jz}}(1-s_{jz}^{2})\frac{\partial F_{j}(s_{jz})}{\partial
s_{jz}}
=-2D\bar{s}_{jz};\: j=1,2;\nonumber\\
\bar{s}_{1,2z}&=& s_{1,2z}(0)e^{-2Dt}.
\label{eq:diff}\end{eqnarray}
 As discussed in \cite{levanpla} (for the case without anisotropy field),
  essential  for the validity of
 this diffusion type  dynamics is that the underlying classical dynamics is
chaotic  in which case the above derivations are
justified.

%
%
\begin{figure}[t]
 \centering
  \includegraphics[width=10cm]{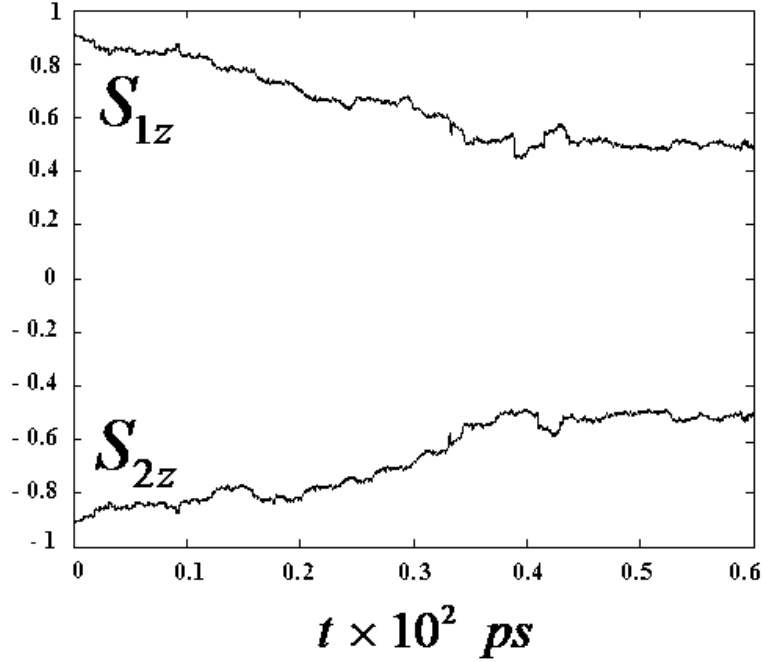}
  \caption{The time evolution of $s_{1z}$ and $s_{2z}$ for the following parameters of the system: $J=0.2$,~~$2\beta=0,1$,~~$H_{0z}=0.2$,~~$\varepsilon=0.057$,~~$\Omega=100$.~~$\frac{\tau_{0}}{T}=0.1$,~~$s_{1z}(0)=0.8$,~~
  $\varphi_{1}(0)=0$,~~$s_{2z}(0)=-0.8$,~~$\varphi_{2}(0)=0$,~~$T=\frac{2\pi}{\Omega}$ and $K_{0}=K'=0.45\cdot10^{-4}$. } \label{Fig:2}
\end{figure}
%
%
%
The stroboscopic evolution of the
spin variables before ($t_{0}-\tau$) and after ($t_{0}+\tau$)
applying the field pulses at  $t= t_{0}$ is expressed as  \cite{levanpla}
\begin{eqnarray}
s_{iz,n+1}&=&s_{iz,n}+\varepsilon T
\sqrt{1-s_{iz,n}^{2}}\sin\varphi_{i,n}, \nonumber \\
\varphi_{i,n+1}&=&\varphi_{i,n}+\bigg(Js_{jz,n+1}+2\beta
s_{iz}-H_{0z}\bigg)T-\varepsilon
T\frac{s_{iz,n}}{\sqrt{1-s_{iz,n}^{2}}}\cos\varphi_{i,n}.
\label{eq:map}
\end{eqnarray}
%
%
The stability of the trajectories is deduced
from the Jacobian matrix \cite{Chirikov} which also set the condition for the chaotic regime as
($<...>_{t}$ means  time average)
\begin{eqnarray}
|\lambda_i|>1,&&  K>0, \label{eq:liapo}\\
\lambda_{1,2}&=&\frac{(2+K)\pm\sqrt{(K+2)^{2}-4}}{2},~~~\lambda_{3,4}=\frac{(2-K)\pm\sqrt{(K-2)^{2}-4}}{2},
\nonumber\\
K&=&K_{0}\sqrt{
(\beta/J)^{2}+(1-2(\beta/J)^{2})s_{1\perp}s_{2\perp}<\cos\varphi_{1}\cos\varphi_{2}>_{t}},~~~
    K_{0}=\varepsilon T^{2}J.
\label{eq:liab}\end{eqnarray}
Hence we can tune to the chaotic regime by varying the  external
fields parameters, the constant of anisotropy $\beta$, and the coupling
constant $J$ between adjacent spins.
 For evaluating averages of the form $<...>_{t}$
  averages over time correlation functions of the random
phases $<\cos\varphi_{1}\cos\varphi_{2}>$ should be considered.
For the correlation term we proceed
as follows: When deriving
the diffusion equation we assumed that correlation time of random
phases are small with respect to the diffusion scale $\tau_{c}<<
\frac{2D}{\pi \varepsilon}= 1/ \Omega $. Taking into account that
in this time scale values $s_{1,2\bot}$ are slow in time, after
averaging of correlation functions over the time interval $\Delta
t\in\left(0,1/\Omega\right)$ we obtain $K=K_{0}\sqrt{
(\beta/J)^{2}+(1-2(\beta/J)^{2})\frac{\varepsilon
\pi}{2D}s_{1\perp}s_{2\perp}\tau_{c}(J)}.$

\subsection{Discussions and numerical results}
Having discussed the analytical structure of the spin dynamics
 we compare the analytical predictions with   full numerical simulations of the problem.
\begin{figure}[t]
 \centering
  \includegraphics[width=10cm]{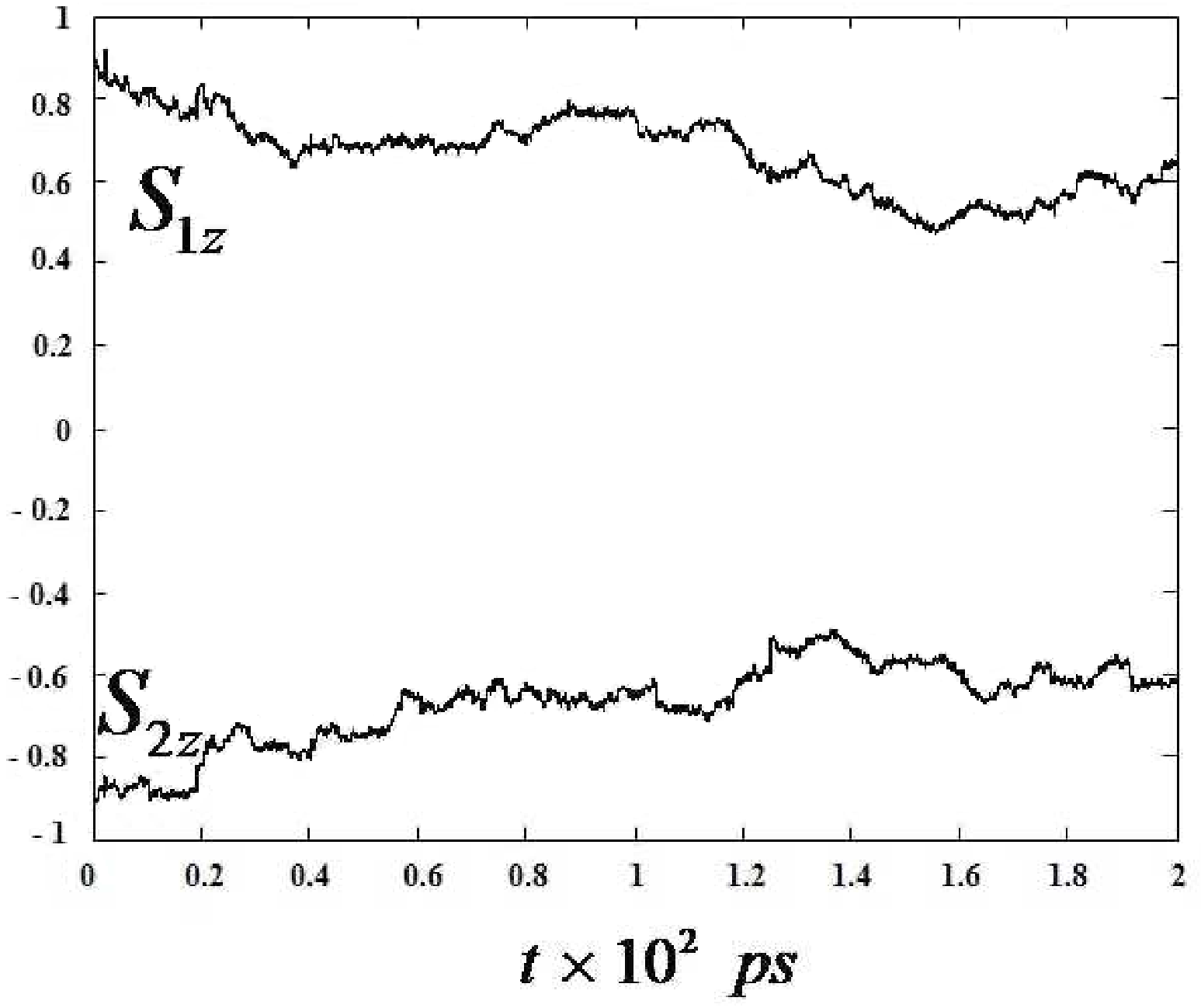}
  \caption{Same quantities as in Fig.(\ref{Fig:2}) however,  $J=0.2$, $2\beta =0.1$, $H_{0z}=0.2$,~~$\varepsilon
=0.04$,~~$\Omega=100$.~~$\frac{\tau_{0}}{T}=0.1$,~~$s_{1z}(0)=0.9$,~~
  $\varphi_{1}(0)=0$,~~$s_{2z}(0)=-0.9$,~~$\varphi_{2}(0)=0$,~~~$T=\frac{2\pi}{\Omega}$ and $K_{0}=0.32\cdot10^{-4}<K'$. } \label{Fig:3}
\end{figure}
\textbf{Note, our system is such that the conditions of chaotic
regime eq.(19),(20) can be realized for arbitrary small
perturbation $\varepsilon>0,~~K_{0}>0$. However, the smallest
values of $\varepsilon$, and corresponding $K_{0}=K^{'}$ that
allows for an observable effect has to be found numerically.}
 At first, the external
 fields are  tuned to  $K_{0}=K'=0.45\cdot10^{-4}>0$. In accord with the analytical results
   stochastic switching of the initial spins direction occur accompanied with a
  subsequent long-time stabilization (see Fig.(2)).
If the  fields are such that
 $K_{0}<K'$ (i.e. $K_{0}$ is very small)  switching does not happen
(see Fig.(\ref{Fig:3})), i.e. $s_{1,2z}$ is still an  adiabatic
invariants;   external fields lead to  small fluctuations
around the equilibrium state. We note that in Fig.(\ref{Fig:3}) the anisotropy field is finite but
its effect is hardly observable becomes of it smallness $\left( {\beta /J} \right)^2  = 0.06$.
The regular (but  non-integrable) regime is reached  by applying very strong
fields ($\varepsilon T\gg J$, $H_{oz}\gg Js^{z}>2\beta s^{z}$)  (cf. eq. (\ref{eq:h})). In this case
no stochastic switching
 occurs (cf. Fig.\ref{Fig:4}.).
  The eigenfrequency of the system is given by the
  constant magnetic field  $\omega_{j}(S_{iz})=JS_{iz}+H_{0z}+2\beta s_{jz}\approx H_{0z}$.
  Physically, effects  related to the exchange interaction and to the anisotropy field become
  negligible and we end up with the familiar resonant switching scheme (this is true  only during the
  time when the  external  fields are on. Effects of exchange and anisotropy govern the subsequent
  field-free dynamics. A scheme for a field-induced deflection and freezing has been proposed
  in Ref.[\onlinecite{prl09}]).
  %
%
\begin{figure}[t]
 \centering
  \includegraphics[width=16cm]{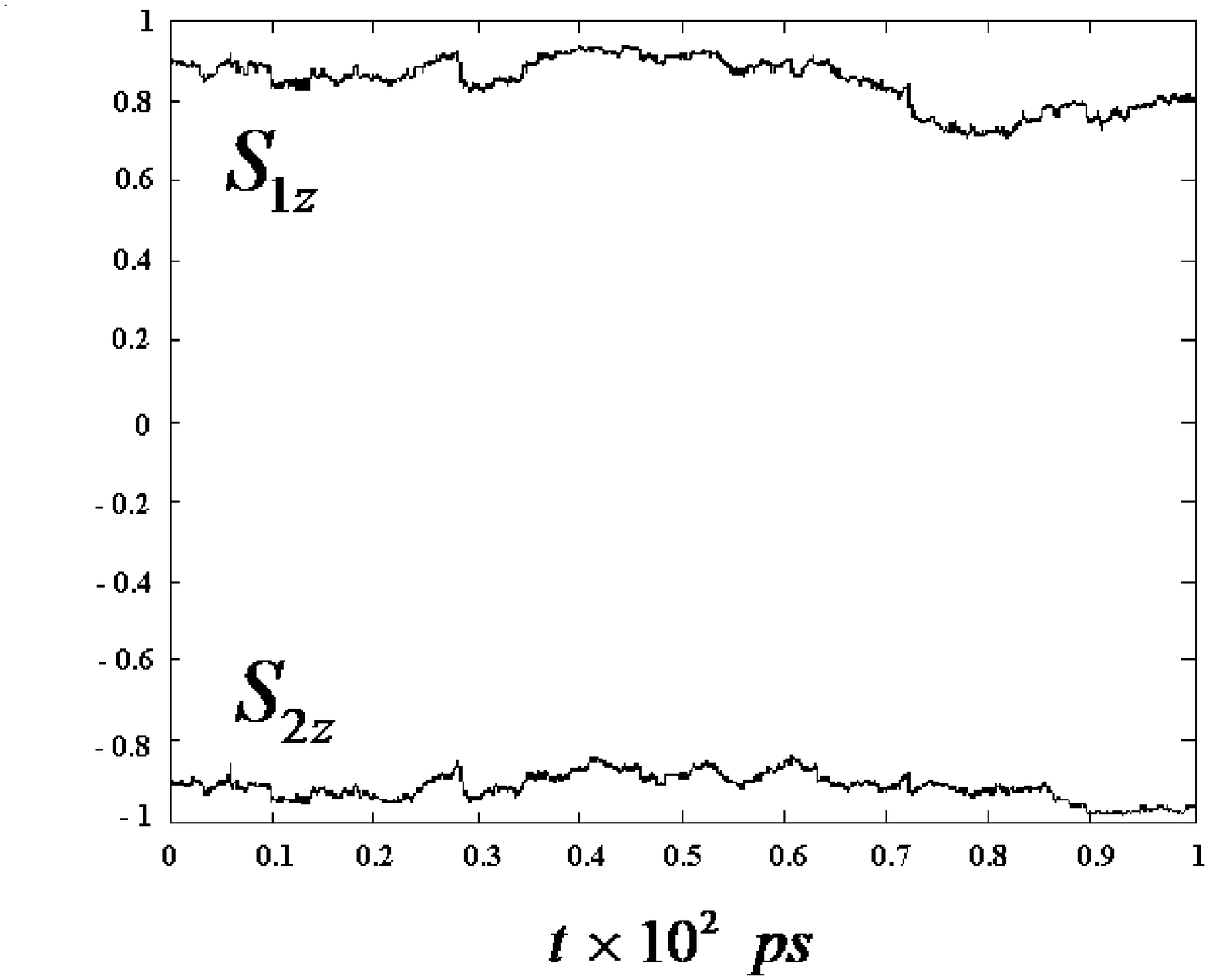}
  \caption{Same quantities as in Fig.\ref{Fig:3} however, $J=0$,~~$\beta=0$,~~$ H_{0z}=0.2$,~~$\varepsilon=0.04$,~~$\Omega=100$.~~$\frac{\tau_{0}}{T}=0.1$,~~$s_{1z}(0)=0.9$,~~
 $\varphi_{1}(0)=0$, ~~$s_{2z}(0)=-0.9$,~~$\varphi_{2}(0)=0$,~~$T=\frac{2\pi}{\Omega}$ and $K_{0}=0$.} \label{Fig:4}
\end{figure}
\subsection{Finite spin chain}
The EOM for a finite spin chain governing the dynamics of each particular
spin follows from
(\ref{eq:ham}) as
$$ \frac{ds_{iz}}{dt}=-\varepsilon\frac{\partial V_{i}\left(\varphi_{i},s_{iz}\right)}{\partial\varphi_{i}},~~~~~~~~~~~~~~~~~~~~ $$
\begin{equation}
\frac{d \varphi_{i}}{dt} =
\omega_{i}\left(s_{i-1z},s_{iz},s_{i+1z}\right)+\varepsilon\frac{\partial
V_{i}\left(\varphi_{i},s_{iz}\right)}{\partial s_{iz}},
\end{equation}
          $$ \omega_{i}\left(s_{i-1z},s_{iz},s_{i+1z}\right)=Js_{i-1z}+Js_{i+1z}+2\beta s_{iz}+H_{0z},~~~i=1,..N.~~~~N+1=N. ~~~~~~~~~~~~~$$
If the variable field has a spatial extent such that only  two spins in the chain, labeled $(k,k+1)$,  are
affected then we find
$$\frac{ds_{iz}}{dt}=-\left(\delta_{i,k}+\delta_{i,k+1}\right)\varepsilon\frac{\partial V_{i}\left(\varphi_{i},s_{iz}\right)}{\partial\varphi_{i}},~~~~~~~~~~~~$$
\begin{eqnarray}
\frac{d \varphi_{i}}{dt} &=&
\omega_{i}\left(s_{i-1z},s_{iz},s_{i+1z}\right)+\left(\delta_{i,k}+\delta_{i,k+1}\right)\varepsilon\frac{\partial
V_{i}\left(\varphi_{i},s_{iz}\right)}{\partial s_{iz}},\nonumber\\
s_{iz}(t)&=&constant \quad \mbox{if}\quad k\neq i \neq k+1.
\label{eq:motion}\end{eqnarray}
\begin{figure}[t]
 \centering
  \includegraphics[width=16cm]{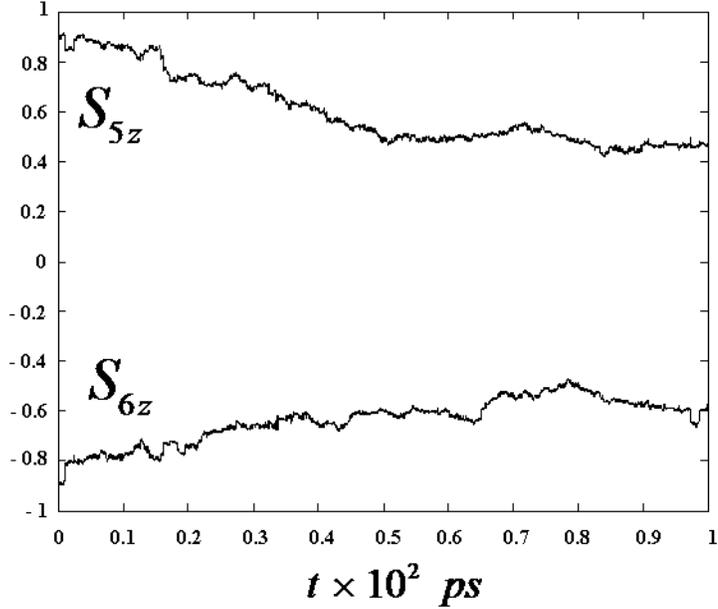}
  \caption{We consider a chain of 10 interacting spins.
  We show here the dynamics of the $z$ components $s_{5,z}$ and $s_{6,z}$ of the spins labelled $5$ and $6$.
   Other parameters are $J=0.2$, ~~ $2\beta=0,3$,~~
  $H_{0z}=0.2$,~~$\varepsilon=0.04$,~~$\Omega=100$.~~$\frac{\tau_{0}}{T}=0.1$,~~$s_{2n+1}^{z}(0)=0.9$,~~
$s_{2n}^{z}(0)=0.9$,~~$\varphi_{i}(0)=0$,~~$T=\frac{2\pi}{\Omega}$,~~$i=\overline{1,N}$.
~~ $N=10$. } \label{Fig:5}
\end{figure}
These equations show that the $z$ component of the spins subjected to the pulse
 have to be determined self-consistently.
 The  dynamics of the oscillation frequency of the spins transverse components
  $\dot{\varphi}_{i},~~i=1...N $
 is determined by  the effective
magnetic field as \be
\dot{\varphi}_{i}(t)=\omega_{i}^{eff}=\gamma_{s}H_{i}^{eff}(t).\ee
Here
$$H^{eff}(t)=\frac{1}{\gamma_{s}}\bigg[H_{0}+J(S_{i-1,z}+S_{i+1,z})+\bigg(2\beta-(\delta_{i,k}+\delta_{i,k+1})\frac{H_{0x}(t)}{S_{i\bot}}\textrm{cos}(\varphi_{i})\bigg)S_{i,z}\bigg].$$
This indicates that the spins subject to the pulses exchange
energy with their nearest neighbors (whose $z$ components are
nevertheless constant). This process depends on the values of the
$z$ components and on the effective frequency
$\omega_{i}^{eff}(t)s_{i}^{z}(t)$; a demonstration of this
phenomena is shown in Fig.\ref{Fig:5}.

\begin{figure}[t]
 \centering
  \includegraphics[width=16cm]{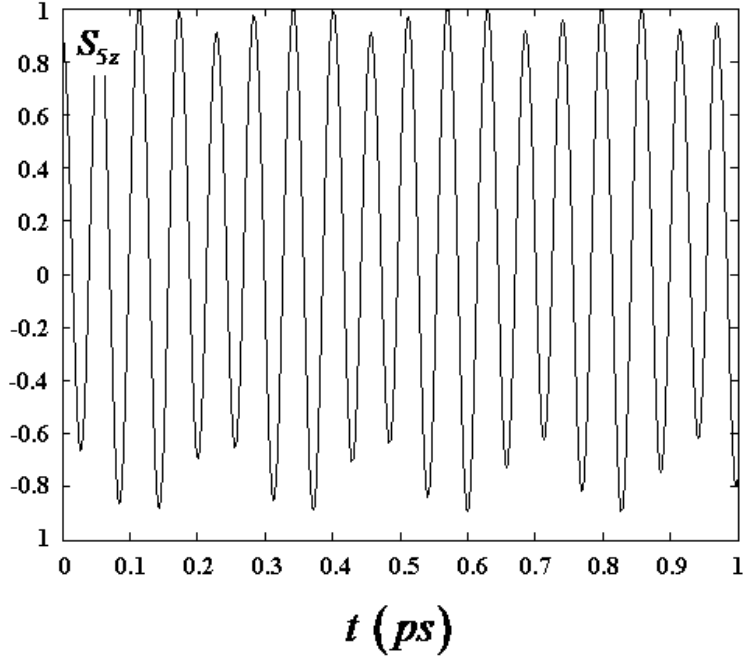}
 \caption{The time dependence of the $z$ component of $5^{rm th}$ spin for the
 parameters
 $J=0.2$,~~$H_{0z}=0.2$,~~$\beta=0$~~$H_{0x}=1$,~~$s_{2n+1}^{z}(0)=0.9$,
 ~~$s_{2n}^{z}(0)=-0.9$,~~$\varphi_{i}(0)=0$,~~$i=\overline{1,N}$,~~$N=10$.}
\label{Fig:6}
\end{figure}

\begin{figure}[t]
 \centering
  \includegraphics[width=16cm]{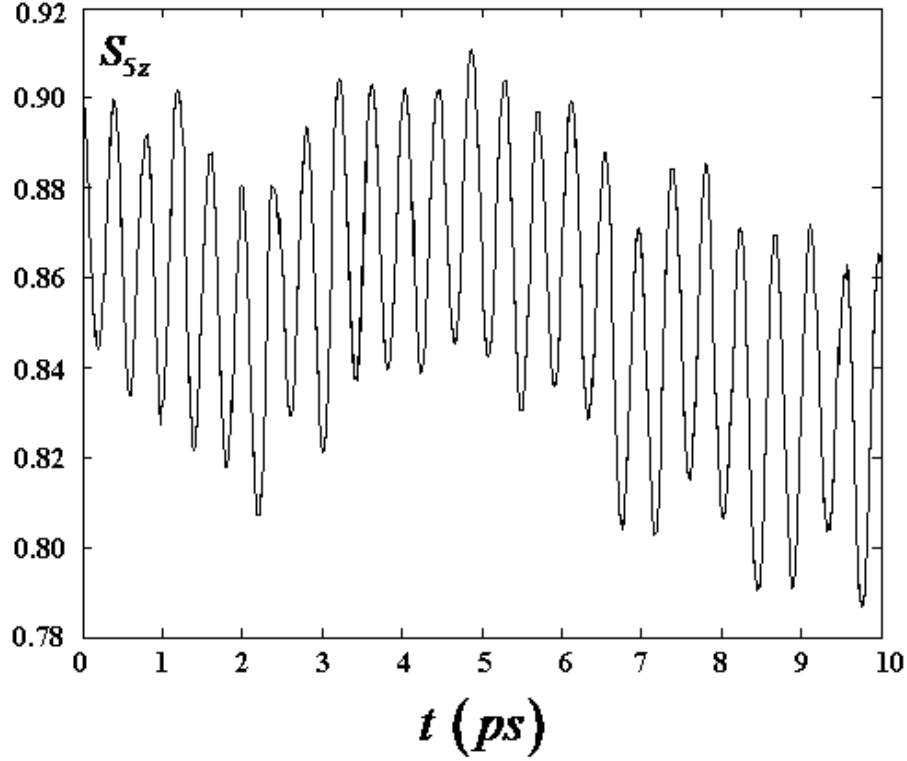}
 \caption{The same as in Fig. \ref{Fig:6} however the parameters are changed to $J=0.2$,~~$H_{0z}=0.2$,~~$H_{0x}=0.01$,
 ~~$s_{2n+1}^{z}(0)=0.9$,~~$s_{2n}^{z}(0)=-0.9$,~~$\varphi_{i}(0)=0$,~
 ~$i=\overline{1,N}$,~~$N=10$.}
\label{Fig:7}
\end{figure}

 A further tool for controlling the
diffusion process is to apply a constant field along the $x$ axis. The system dynamics
 is then chaotic, even without the periodic series of
pulses \cite{Mejia-Monasterio,Saito}. Therefore, the $z$ component of the
spin is not an adiabatic invariant and the mechanism of dynamical freezing (DF) discussed above
 does not work.
Fig.\ref{Fig:6} illustrates that if the amplitude of the magnetic field
applied along $x$ axis is strong enough
$H_{0x}>H_{0z}$ then the longitudinal component of the spin performs fast
oscillations. In the other opposite case $H_{0x}<H_{0z}$ the orientation
of the spin can be deflected but DF again is not possible
(cf. Fig \ref{Fig:7}).
\begin{figure}[t]
 \centering
  \includegraphics[width=16cm]{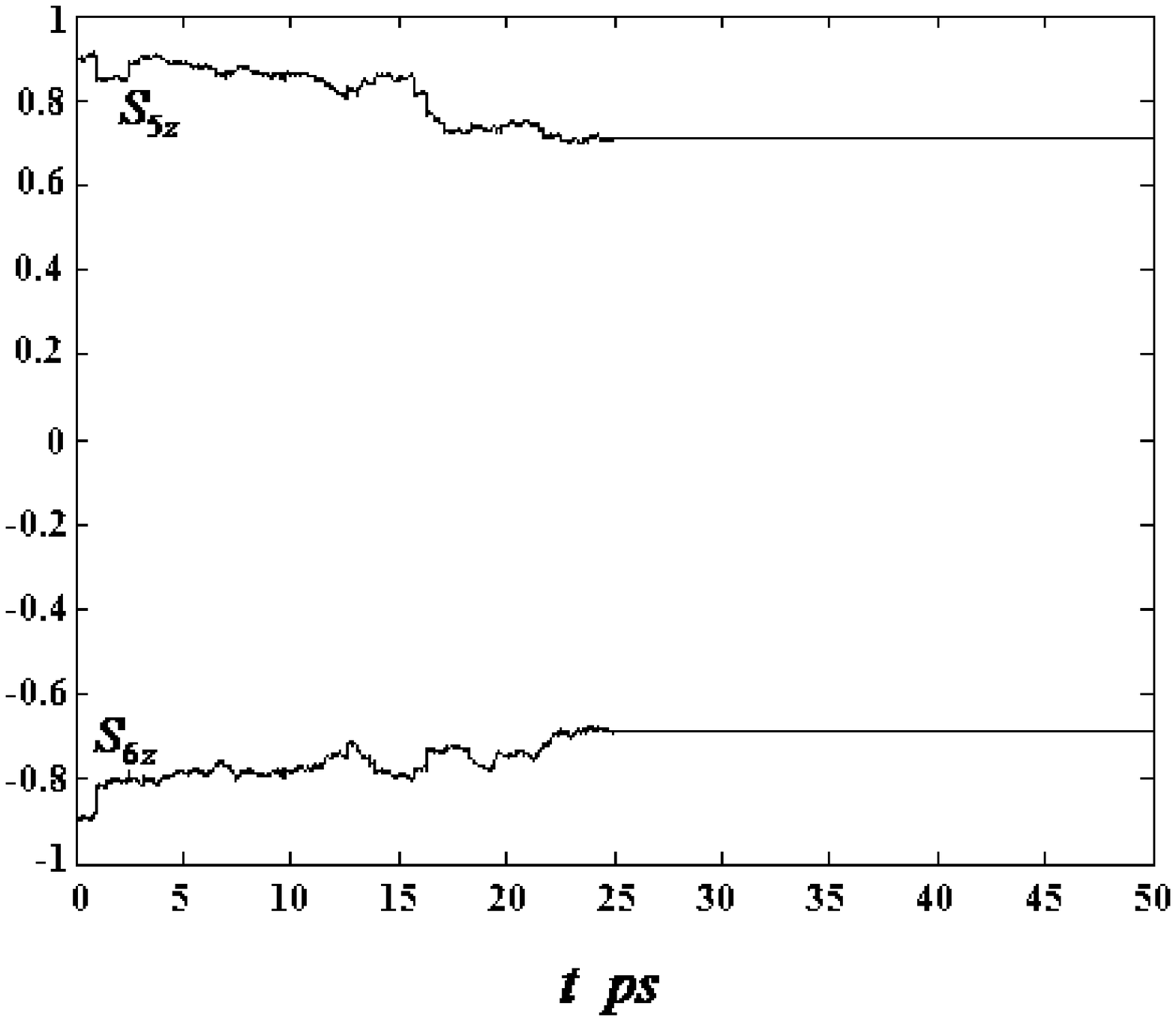}
 \caption{ The $z$ components of the spins labelled $5$ and $6$
 for the parameters: $J=0.2$,~~$H_{0z}=0.2$,~~$H_{0x}=0.01$,~~$s_{2n+1}^{z}(0)=0.9$,
 ~~$s_{2n}^{z}(0)=-0.9$,~~$\varphi_{i}(0)=0$,~~$i=\overline{1,N}$,~~$N=10$.}
\label{Fig:8}
\end{figure}

After deflection of  the spin to a desired angle, one can completely
freeze its orientation.  Key point is the fact that, in the
absence of pulses the $z$ component of the spin projection is an
integral of motion. The system of equations in this case has the form
\begin{eqnarray}\label{eq:dis}
&&
\frac{ds_{jx}}{dt}=-J\big(s_{j-1,z}+s_{j+1,z}\big)s_{jy}-H_{0z}s_{jy},\nonumber\\
&& \frac{ds_{jy}}{dt}=J\big(s_{j-1,z}+s_{j+1,z}\big)s_{jx}+H_{0z}s_{jx},\\
&& \frac{ds_{jz}}{dt}=0. \nonumber
\end{eqnarray}
For simplicity we assumed  here that $\beta=0$. After solving(24)
one  obtains
\begin{eqnarray}\label{eq:sol}
&&
s_{jx}=s_{jx}\big(t_{0}\big)\cos\big(\omega_{j0}t+\varphi_{0}\big),\nonumber\\
&& s_{jy}=s_{jy}\big(t_{0}\big)\sin\big(\omega_{j0}t+\varphi_{0}\big),\\
&& s_{jz}=s_{jz}\big(t_{0}\big), \nonumber
\end{eqnarray}
where
$\omega_{j}\bigg(s_{j-1z}\big(0\big);s_{j+1z}\big(0\big)\bigg)=J\bigg(s_{j-1z}\big(0\big)+s_{j+1z}\big(0\big)\bigg)+H_{0z}=\omega_{j0}$,
and $s_{jz}\big(t_{0}\big)$  corresponds to the desired
orientation of the spin, achieved after the action of the pulses in the time
interval $0<t<t_{0}$ (cf Fig.\ref{Fig:8}).
%

\subsection{Arnold Diffusion}

Results of previous section evidence that even in the case of a long
spin chain  the orientation of  spins can be still  controlled. This follows
from resonance overlapping and the existence of diffusion.
However, the question of what kind of diffusion we have is still outstanding.
 If the dimension
of the system is more than $N>2$, the dynamics is much more
involved    and the emergence of new physical phenomenon is expected.
We recall the key idea of KAM theory: The size of the
destroyed torus is small and the domain of their location is
surrounded by invariant  torus. This situation changes
if an invariant torus crosses the domain of the destroyed torus
location. This is possible if and only if $N>2$.
The phenomenon of universal diffusion along the net formed due to the
inter-tours crossing was discovered by Arnold \cite{Sagdeev}. Here we consider the
mechanism of the formation of the stochastic net in the case of a spin
chain.
\begin{equation}
H=H_{0}(s^{z}_{1},...s^{z}_{N})+\varepsilon V(\varphi_{1},.....\varphi_{N}).
\end{equation}

Note,  the frequencies  of  the unperturbed motion on the $N$ dimensional
torus is a function of the three actions
\begin{equation}
\omega_{i}(s^{z}_{i-1}, s^{z}_{i}, s^{z}_{i+1})= Js^{z}_{i-1}+Js^{z}_{i+1}+2 \beta s^{z}_{i}+H_{0z},~~~~~i = \overline{1,N}.
\label{eq:om}\end{equation}
We collect the resonant tours defined by the condition:
\begin{equation}
\sum\limits_{j=1}^{N}n_{j}\omega_{j}=0,
\label{eq:n}\end{equation}

where $n_{j}$ are  integer numbers. For each set of numbers there
exists a multitude of solutions
$s^{0}_{z}\equiv(s^{z(0)}_{1},...s^{z(0)}_{N})$. Each solution
determines resonant tours.  For the formation of the Arnold diffusion
the absences of degeneracy is essential
\begin{equation}
det\bigg|\frac{\partial^{2} H_{0}}{\partial s^{z}_{i}\partial s^{z}_{j}}\bigg|\neq0, ~~~~i,j=\overline{1,N}.
\label{deg}\end{equation}
In the case of our system due to the form of the matrix
\begin{equation}
\frac{\partial^{2} H_{0}}{\partial s^{z}_{i}\partial s^{z}_{j}}= \left( \begin{array}{c}
  2\beta~~J~~0~~0~~. \\
J~~2\beta~~J~~0~~. \\
0~~J~~2\beta~~J~~. \\
0~~0~~J~~2\beta~~. \\
  .~~.~~.~~.~~. \end{array} \right)
,\end{equation}
the condition of the absence of a degeneration (29) leads to the
polynomial expression
\begin{equation}
det\bigg|\frac{\partial^{2} H_{0}}{\partial s^{z}_{i}\partial s^{z}_{j}}\bigg|=G(J, \beta, N)\neq0 .
\label{eq:g}\end{equation}
The explicit form of the expression (\ref{eq:g}) also depends on
the system's size (in addition to the dependence on the parameters
$\beta, J $).  For
 large systems we obtain the following asymptotic  expressions
\begin{equation}
G(J, \beta, N)=J^{N} ~~if~~ J>\beta,~~~~G(J, \beta, N)=2^{N}\beta^{N} ~~if~~ J<\beta .
\end{equation}
From this relation we can conclude that the universal diffusion is
possible for any nonzero $J, \beta$, and identify the numeric
results obtained for the spin chain with the Arnold diffusion. For
an analytical estimation we consider the minimal possible
dimension.  Therefore,  in what follows, without loss of
generality we shall restrict ourselves by the case $N=3$. From
eq.(28) we find
\begin{equation}
n_{1}\omega_{1}+n_{2}\omega_{2}+n_{3}\omega_{3}=0 ,
\label{eq:n3}\end{equation}
where each frequency depends on the three action according to
(27). In the frequency space $(\omega_{1},\omega_{2},\omega_{3})$
eq. (33) determines a family of surfaces. On the energy surface we
have
\begin{equation}
H_{0}(s^{z}_{1},s^{z}_{2},s^{z}_{3})=E .
\label{eq:h0}\end{equation}
This equation  is also the equation determining the surface (33).
Therefore, the resonant tours have a common parts along the
curves, defined as  the solutions of the set of equations (33),
(34). The time dependent perturbation leads to a widening of these
curves and to the formation of the stochastic net.

In order to provide topological interpretation of this phenomenon
we will consider simplest case of three spins. In this case the
explicit form of Eq (33) and Eq (34) reads:

\be
(J+2\beta)(\omega_{1}^{2}+\omega_{2}^{2}+\omega_{3}^{2})-2\omega_{1}\omega_{2}-2\omega_{1}\omega_{3}-2\omega_{2}\omega_{3}=E,\ee
and \be n_{1}\omega_{1}+n_{2}\omega_{2}+n_{3}\omega_{3}=0.\ee

Here $E=(8\beta^{2}+4\beta J-4J^{2})H_{0}-3H_{0z}^{2}(2\beta-J),$
is a re-scaled energy. From the equation (36) one can exclude
frequency $\omega _1  =  - \frac{{n_2 }}{{n_1 }}\omega _2 -
\frac{{n_3 }}{{n_1 }}\omega _3 $ and rewrite the  equation (35) as
a function of the two frequencies $\left( {\omega _3 ,\omega _2 }
\right)$ . Clearly,  the shape of the implicit plot for $\omega _3
\left( {\omega _2 } \right)$ depends on the values of parameters
$\left( {n_1 ,n_2 ,n_3 } \right)$. Therefore, we expect that the
$\omega _3 \left( {\omega _2 } \right)$, plotted for different
inner resonances $\left( {n_1 ,n_2 ,n_3 } \right)$ should cross at
some points. Now one can construct implicit plots expressing
frequencies as a function of each others for different resonances
(cf. Fig.9). We see that, in some points trajectories cross each
other. Due to topological reasons, such a nodal points are
possible if and only if systems dimension is at least $N=3$ or
higher. Nodal points are crossing points between different
resonances. If an external adiabatic perturbation is applied the
dynamic near the nodal points becomes unpredictable and this leads
to the Arnold diffusion [6].

\begin{figure}[t]
 \centering
  \includegraphics[width=6cm]{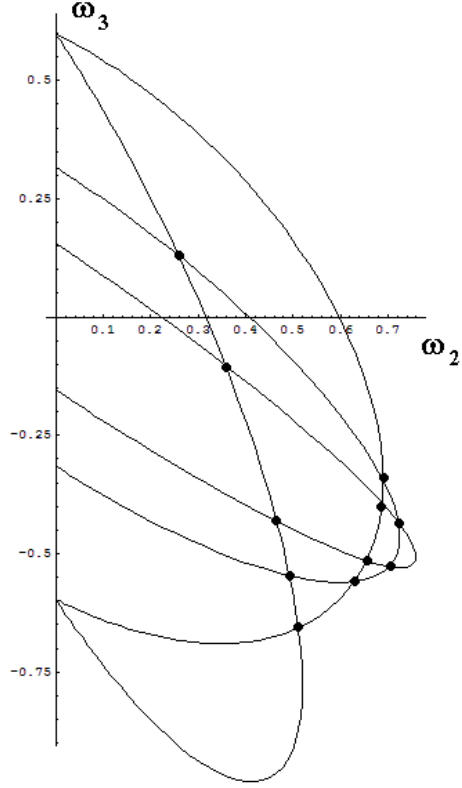}
 \caption{Topological structure of the inner resonances on the frequency plane $(\omega_{3},\omega_{2})$ plotted for different resonances:
 $(n_{2}=n_{1},n_{3}=n_{1}),~~~(n_{2}=2n_{1},n_{3}=3n_{1}),
 ~~~(n_{2}=3n_{1},n_{3}=n_{1}),~~~(n_{2}=5n_{1},n_{3}=7n_{1}).$ Further values: $J = 0.2,\,\,\,\,\,\,2\beta  = 0.3$.
}
\label{Fig:9}
\end{figure}

In the general  dimensional  case ($N>>1$), the geometrical
interpretation is less illustrative and much more complicated.
Since we have to deal with $N-1$ dimensional  hyper-curves in the
N dimensional hyper-space, the basic concept is however the same
[20]). This conclusion manifests fundamental  features  of the
multidimensional nonlinear dynamical systems. The diffusive motion
of the system in the stochastic net is named as Arnold diffusion.
Therefore, the diffusion equation (16) is still  justified.
However, the coefficient of diffusion for Arnold diffusion is
defined by an expression other than eq. (16), namely
\cite{Sagdeev}:
\begin{equation}
D_{A}=E^{2}\, \varepsilon\, H_{0z}\, e^{-{1}/{\varepsilon^{a(N)}}}.
\end{equation}
Here $E$ is the system's energy, $a(N)$ is a dimensional dependent
scaling constant with an upper limit determined by the Arnold
inequality relation [6]
\begin{equation}
a(N)<\frac{2}{6N(N-1)+3N+14}.
\end{equation}
Obviously, for $N>>1$ , $a(N)\mapsto0$ and the coefficient of the
Arnold diffusion takes the more simple form
\begin{equation}
D_{A}=\frac{1}{e}\varepsilon E^{2}H_{0z}.
\end{equation}
Comparing  eq. (39) with the diffusion coefficient obtained for
the case of two spins, i.e. eq. (16) we find
\begin{equation}
\frac{D_{A}}{D}=\frac{4E^{2}H_{0z}}{e\varepsilon T}.
\end{equation}

This  relation is important in that it delivers information on when the
mechanism of stochastic switching and dynamical freezing are
more efficient $\frac{D_{A}}{D}>1$ for a long spin chain, as compared
to the case of a pair of spins.
\subsection{Role of anisotropy field}
Here we discuss the connection between the anisotropy field and Arnold diffusion.
For the Arnold diffusion to occur the Jacobi matrix has to be none-degenerate. We note however, that the
Jacobi matrix becomes degenerate in some cases if the anisotropy field is absent, as can be inferred from the
 structure of the Jacobi matrix.  For example in the simplest case of three spins  ,
$$\det \left| {\frac{{\partial ^2 H_0 }}{{\partial S_i^z \partial S_j^z }}} \right| =  - 4J^2 \beta  + 8\beta ^3  \ne 0,\,\,\,\,\,\,\,\,\,\,\,\,\,if\,\,\,\,\,\,\,\,\,\beta  \ne 0.$$
Evaluating the determinant for different number of spins we find  that with the anisotropy field  being applied it is always non-degenerated, while for particular $N$, it becomes degenerated in absence of the anisotropy field.
\section{Low temperature limit}
\begin{figure}[t]
  \centering
  \includegraphics[width=16cm]{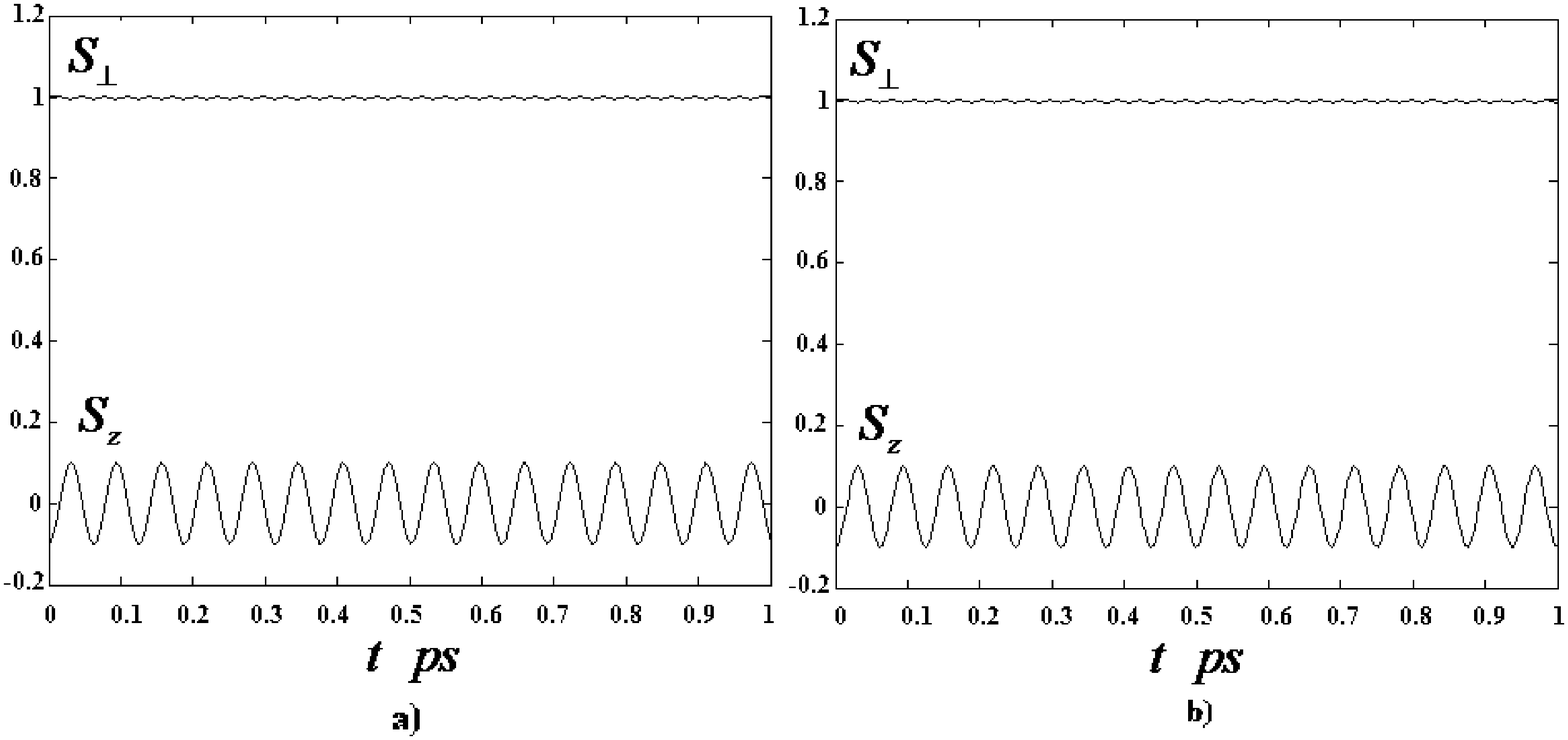}
  \caption{The longitudinal $s_{z}$, and the transversal
  $s_{\perp}=\sqrt{s_{x}^{2}+s_{y}^{2}}$   spin components as a function of time.
    $J=0.05$ ,~~$H_{0z}=1$, $H_{0x}=0.1$. Graph a) shows the  analytical solutions given by eq.(52), whereas
     in graph b) the numerical integration of the system of equations (\ref{eq:cont}) is depicted.}
\label{Fig:11}
\end{figure}
\begin{figure}[t]
  \centering
  \includegraphics[width=16cm]{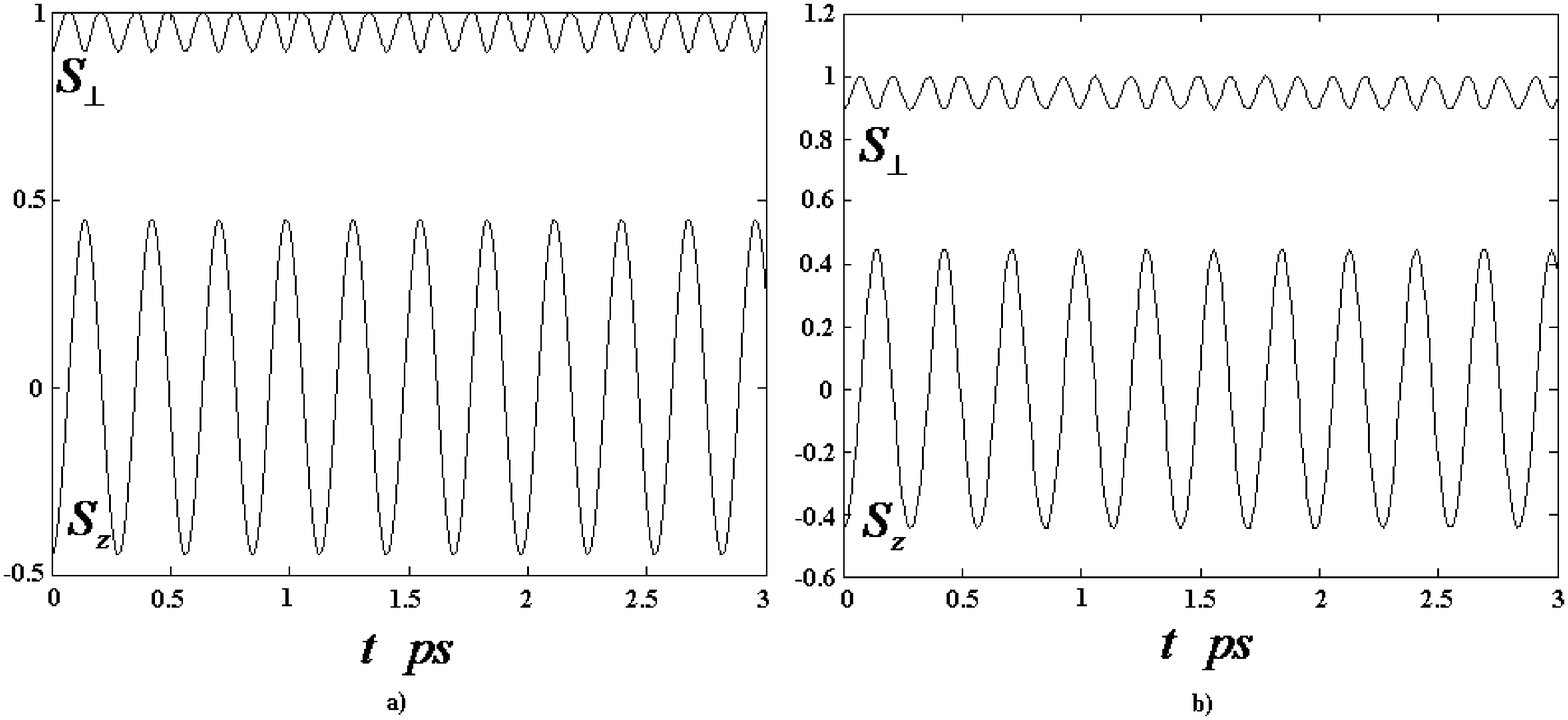}
  \caption{The same as in Fig.10 with the same meaning of the labels. The parameters of the
  system are however changed to
    $J=0.05$ ,$H_{0z}=0.2$, $H_{0x}=0.1$.}
\label{Fig:12}
\end{figure}
In this section we consider the dynamics in the continuous  limit.
 By a proper choice of pulse parameters we were able to
 deflect the spin orientation diffusively  to a desired angle. Switching off the
pulses, the dynamics remain quasi  frozen  (is equivalent to the spin $z$ component being
an integral of motion).
 The question we pose here is that what happens if upon stochastic switching and freezing
  we apply  a constant magnetic field along $x$ axis.
 We recall that applying a constant field  to the equilibrium (initial) state
  along the $x$ axis invalidates the use of the KAM theory and the
  mechanism of SC does not work (cf.eq.(\ref{eq:conv})).
  Before we deal with this problem in more details
  we set the limits of continuous approximation.
   Due to the constant magnetic field, applied along  the $x$
 axis, the $s_{z}$ component is not an integral of motion any more.
 Therefore, excitations similar to spin waves  propagate along the  spin chain.
 These waves are not completely  analogous to   spin waves because  the non-conservation
 of $s_{z}(t)$  is  not related to flip-flop processes, but requires a transversal magnetic field.
 However, the wavelength of such excitations can be evaluated in a manner similar to the spin
 waves  case \cite{Kittel}.
  If the wavelength is larger than the distance between the spins $\lambda>>a$ a continuous treatment is justified.
Taking into account the expression for the wave frequency
 \begin{equation}
 \omega=\frac{4|J|S}{\hbar}\cdot\frac{2\pi}{\lambda}\cdot a,
 \end{equation}
One  concludes that the  validity of the continuous  approximation
depends on the temperature
\begin{equation}
 T<<\frac{J\hbar}{k_{B}}.
 \label{eq:TJ}
 \end{equation}
 Here $k_B$ is the Boltzmann constant.
Thus, the continuous approximation corresponds to a low temperature approximation.
  For anti-ferromagnetic materials FeCl$_{2}$, or CoCl$_{2}$  we have $J=1.23\cdot10^{12}$ Hz,
    from (42)  we infer for temperature regime of the continuous model
  $T<3 K$.

  Returning back to the spin chain in a static magnetic field along $x$ axis, the EOM read
  \begin{eqnarray}\label{eq:disc}
&&
\frac{ds_{jx}}{dt}=-J\big(s_{j-1,z}+s_{j+1,z}\big)s_{jy}-H_{0z}s_{jy},\nonumber\\
&& \frac{ds_{jy}}{dt}=J\big(s_{j-1,z}+s_{j+1,z}\big)s_{jx}+H_{0z}s_{jx}-H_{0x}s_{jz},\\
&& \frac{ds_{jz}}{dt}=H_{0x}s_{jy}. \nonumber
\end{eqnarray}
Considering that
 \begin{eqnarray}
&&
s_{jx}\rightarrow s_{x}(x;t),\nonumber\\
&& s_{jy}\rightarrow s_{y}(x;t),\\
&&s_{jz}\rightarrow s_{z}(x;t) \nonumber
\end{eqnarray}
\begin{eqnarray}
&&
s_{j-1;z}=s_{x}(x,t)-a\frac{\partial s_{z}(x,t)}{\partial x}+\frac{a^{2}}{2}\frac{\partial^{2}s_{x,t}}{\partial t^{2}},\nonumber\\
&& s_{j+1;z}=s_{x}(x,t)+a\frac{\partial s_{z}(x,t)}{\partial
x}+\frac{a^{2}}{2}\frac{\partial^{2}s_{x,t}}{\partial t^{2}},
\end{eqnarray}
from (43) we deduce that
\begin{eqnarray} \label{eq:con}
&&
\frac{\partial s_{x}}{\partial t}=-Js_{y}\bigg(2s_{z}+\frac{\partial^{2}s_{z}}{\partial x^{2}}a^{2}\bigg)-H_{0z}s_{y},\nonumber\\
&& \frac{\partial s_{x}}{\partial t}=Js_{x}\bigg(2s_{z}+\frac{\partial^{2}s_{z}}{\partial x^{2}}a^{2}\bigg)-H_{0x}s_{z}+H_{0z}s_{x},\\
&&\frac{\partial s_{z}}{\partial t}=H_{0x}s_{y}. \nonumber
\end{eqnarray}
In the low temperature regime  we can neglect quadratic
terms in (46) and  obtain
\begin{eqnarray} \label{eq:cont}
&&\dot{x}=-\delta y-\gamma y z, \nonumber \\
&&\dot{y}=\delta x-z+\gamma x z,\\
&&\dot{z}=y. \nonumber
\end{eqnarray}
Here we  introduced the following notations
$$\delta=\frac{H_{0z}}{H_{0x}}, ~~ \gamma = \frac{2J}{H_{0x}},~~ t\rightarrow H_{0x}t, ~~s_{x}=x,~~s_{y}=y,~~s_{z}=z.$$
\textbf{ Eq.(47) is derived in the absence of anisotropy field.
However for the role of the anisotropy field we remark the
following: The Zeeman field applied along the  $z$ axis is very
strong $H_0^z
> J\left| {S_{i - 1}^z } \right|\, + J\left| {S_{i + 1}^z }
\right| + 2\beta \left| {S_i^z } \right|$. The eigenfrequency is
constant and we have no effect of a dynamical shift $\omega _i
\left( {S_{i - 1}^z \,,S_{i + 1}^z ,\,\,S_i^z } \right) = JS_{i -
1}^z \, + JS_{i + 1}^z  + 2\beta S_i^z  + H_0^z \approx H_0^z $.
Inclusion of a finite anisotropy field leads to a rescaling of the
small parameter $\gamma$ in in Eq. (47), i.e. $\gamma  =
\frac{{2J}}{{H_{0x} }} \to \frac{{2J + 2\beta }}{{H_{0x} }}$.
Hence, we conclude that in this particular case the anisotropy
field has no principal dynamical effect. Eq.(47) with rescaled
parameter $\gamma =\frac{2J+2\beta}{H_{0x}}$ is still valid in the
presence of anisotropy field.}

When solving (47), we assume for the initial values the spin
orientations achieved after action of pulses. In order to obtain
analytical solutions we utilize the canonical perturbation theory
(cf., e.g. \cite{Ugulava}). The parameter $\gamma$ is assumed to
be small.
  First step is to rewrite (47) in a canonical form.
This can be done using the following transformation. (For details, see
Appendix)
\begin{eqnarray}
&&x_{1}=2\delta x-2z, \nonumber \\
&&y_{1}=2\lambda y,\\
&&z_{1}=\frac{1}{\delta}x+z, \nonumber \\
&&\lambda=\sqrt{1+\delta^{2}}, \nonumber
\end{eqnarray}
Equations (47) assume then the form
\begin{eqnarray} \label{eq:cont1}
&&\dot{x_{1}}=-\lambda y_{1}+\frac{\delta\gamma}{2\lambda^{3}}y_{1}x_{1}-\frac{\gamma\delta^{3}}{\lambda^{3}}y_{1}z_{1}, \nonumber \\
&&\dot{y_{1}}=\lambda x_{1}+\frac{\gamma\delta(\delta^{2}-1)}{\lambda^{3}}x_{1}z_{1}-\frac{\gamma\delta}{2\lambda^{3}}x_{1}^{2}+\frac{2\gamma\delta^{3}}{\lambda^{3}}z_{1}^{2},\\
&&
\dot{z_{1}}=\frac{\gamma}{4\lambda^{3}\delta}y_{1}x_{1}-\frac{\gamma\delta}{2\lambda^{3}}y_{1}z_{1} . \nonumber
\end{eqnarray}
 We seek a solution of (49) having the structure
 \begin{eqnarray}
 x_{1}=Cx_{1}^{1}+C^{2}x_{1}^{(2)}+C^{3}x_{1}^{(3)}+C^{4}x_{1}^{(4)}\ldots,\nonumber \\
 y_{1}=Cy_{1}^{1}+C^{2}y_{1}^{(2)}+C^{3}y_{1}^{(3)}+C^{4}y_{1}^{(4)}\ldots, \\
 z_{1}=Cz_{1}^{1}+C^{2}z_{1}^{(2)}+C^{3}z_{1}^{(3)}+C^{4}z_{1}^{(4)}\ldots,\nonumber
 \end{eqnarray}
 and in addition we use the re-scaled time
 \begin{equation}
 t=\frac{\tau}{\lambda}\big(1+h_{2}C^{2}+h_{3}C^{3}+\ldots\big).
 \end{equation}
 With an accuracy up to  third order in $\gamma$ we find the solution of (47) to be
 \begin{eqnarray}\label{eq:sol}
 && s_{x}=\frac{\delta\cos
 \tau}{\lambda}+\frac{\gamma(1+2\delta^{2})\sin^{2}(\tau)}{2\lambda^{4}}-\frac{\gamma^{2}\delta(9+4\delta^{2})\cos(\tau)\sin^{2}(\tau)}{16\lambda^{7}},\nonumber
 \\
 && s_{y}=\sin(\tau)-\frac{\gamma\delta\cos(\tau)\sin(\tau)}{\lambda^{3}}+\frac{\gamma^{2}(1+4\delta^{2})(-5\sin(\tau)+3\sin(3\tau))}{64\lambda^{6}},\\
&&s_{z}=-\frac{\cos\tau}{\lambda}-\frac{\gamma\delta\sin^{2}(\tau)}{2\lambda^{4}}-\frac{\gamma^{2}(-1+4\delta^{2})\cos(\tau)\sin^{2}(\tau)}{16\lambda^{7}}.\nonumber
 \end{eqnarray}
 Fig.10 demonstrate that these analytical solutions are in a good agreement with the exact numerical simulation of the system (\ref{eq:cont}).
 Fig. 10 evidences that  a constant magnetic field results in oscillations of spin's longitudinal
 component in a controlled manner. If the  amplitude of the magnetic field is small, nonlinear effects become more important
 (cf Fig.11).

\section{Quantum Mechanical consideration and problem of DF}

\begin{figure}[t]
  \centering
  \includegraphics[width=10cm]{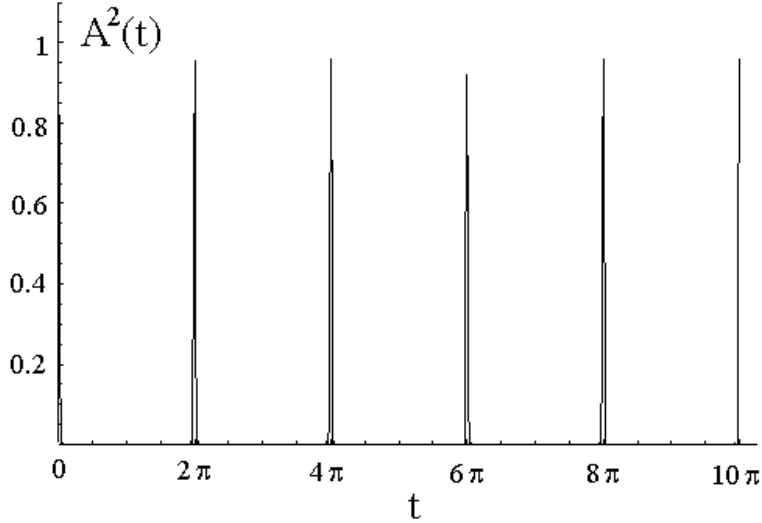}
  \caption{Bargmann  angle $\theta_{B}(t)$
   as a function of re-scaled time $t\equiv t/(H_{0z}+J)$, $H_{0z}=J=0,2$ calculated from (59).}
\label{Fig:13}
\end{figure}

\begin{figure}[t]
  \centering
  \includegraphics[width=16cm]{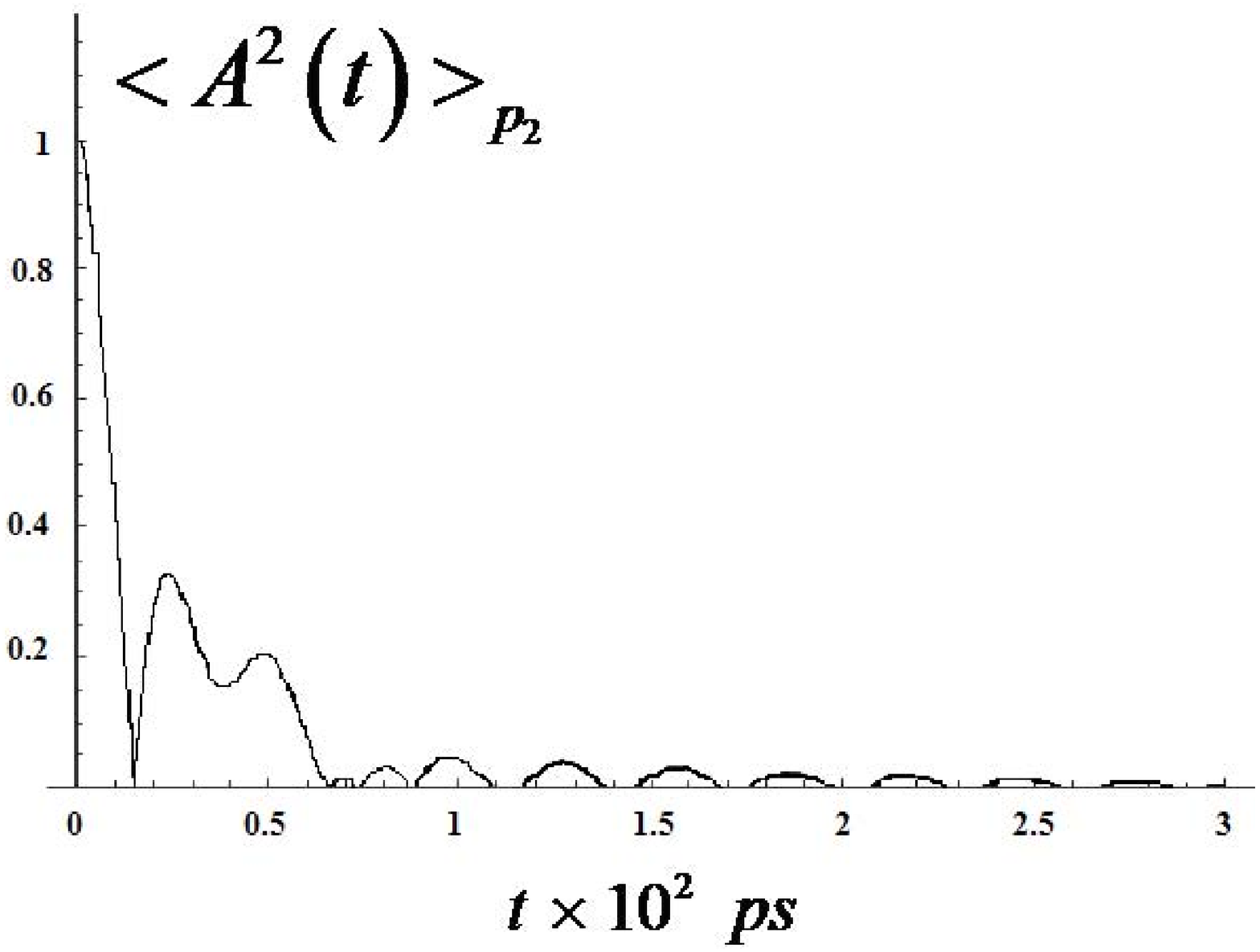}
  \caption{The time evolution of the Bargmann  angle $\theta_{B}(t)$, as
   calculated using eq.(62) and averaged over  the random quantum phases.
    The parameters of the driving fields and the spin chains
    are the same as in Fig.\ref{Fig:2}, i.e. we are in the stochastic switching regime.}
\label{Fig:14}
\end{figure}
As stated above, the classical analysis is useful if the
 atoms in the chain have a large magnetic moment. In fact,
 for a finite chain of manganese (Mn) atoms  \cite{science}  the classical approach proved to be
 adequate \cite{spintheory}. This situation changes however, for small spins where
 the dynamics becomes dominated by quantum effects.
  What is needed for our quantum consideration is the
structure of the energy spectrum in the regime
where the underlying classical dynamics is chaotic \cite{Mejia-Monasterio}.
The point of interest here is that whether quantum effects invalidate
the SC and in particular die dynamical freezing. To this end we use the concept
  of quantum geometry
\cite{Pati,vonBalk} in the way done in Refs.[\onlinecite{Matos-Abiague,Matos-Abiague2}] to study
DF. Let us consider two quantum states
     $\Psi_{1}$  and $e^{i\varphi}\Psi_{1}$.
    The distance in Hilbert space between them
can be characterized by the quantity  \cite{Pati,Anandan,vonBalk}
$$
D_{1}(\Psi_{1},\Psi_{2})=min_{\varphi}\big
\|\Psi_{1}-e^{i\varphi}\Psi_{2}\big \|.
$$
The minimal phase $\varphi_{m}$  is found by exterimizing
$\big\|\Psi_{1}-e^{i\varphi}\Psi_{2}\big\|$  and noting that
$\big\|\Psi_{1}-\Psi_{2}\big\|=\big\langle\Psi_{1}\big|\Psi_{2}\big\rangle^{1/2},$
which yields
\begin{equation}
\exp\big(i\varphi_{m}\big)=\frac{\big\langle\Psi_{1}\big|
\Psi_{2}\big\rangle}{\big|\big\langle\Psi_{1}\big|\Psi_{2}\big\rangle\big|}.
\end{equation}
Therefore, we writes for the  distance  $D_{1}$
$$D_{1}(\Psi_{1},\Psi_{2})=\sqrt{2-2\big|\big\langle\Psi_{1}\big|\Psi_{2}\big\rangle\big|}.$$
For the same purpose as for $D_1$, one may also use the Fubini-Study metric \cite{Anandan}
 \begin{equation}
D_{2}^{2}\big(\Psi(t),\Psi(0)\big)=4\big(1-\big|\big\langle\Psi(t)\big|\Psi(0)\big\rangle\big|^{2}\big).
 \end{equation}
In both approaches the key quantity is  the so-called Bargmann angle
\cite{Matos-Abiague}
\begin{equation}
\cos\theta_{B}(t)=A(t)=\big|\big\langle\Psi(0)\big|\Psi(t)\big\rangle\big|.
\end{equation}
In analogy to the classical deflection in
 quantum control problems, the Bargmann angle  can be considered as the "quantum
deflection".
 Essential for further progress is our (classical) finding that
 SC and DF occur in the classical  chaotic regime. This calls for the use
 of  random matrix theory (RMT) to inspect the quantum dynamics
\cite{Stockman,Haake}.
 To this end we write  (\ref{eq:ham}) as
%
\begin{equation}
\hat{H}(t)=\hat{H}_{0}+\hat{V}(t),
 \end{equation} where
$\hat{H}_{0}$ is  time independent.
Now  we employ the established Floquet-operator method \cite{Haake}
 and the quantum map  and infer  for  the
Bargmann angle
\begin{equation}
A^{2}(t)=\frac{1}{N^{2}}\bigg(N+\sum\limits_{n,m=1\atop n\neq
m}^{N}\cos\big[t(\varphi_{n}-\varphi_{m})\big]\bigg).
\label{eq:Berg}
\end{equation}
Here $\varphi_{n}$
stand for the eigen phases of the Floquet operators.
$N$ is the Hilbert space dimension.
From this relation it is evident that starting at
$t=0$ with two completely coherent  states, i.e.
$A(t=0)=1$, de-coherence sets in for $t\neq0$. To put the classical
predictions of the previous section into a quantum perspective we
note the following: In the classical regular regime, as identified above,
the time-dependent perturbation $\hat{V}(t)$ acts adiabatically and  does not alter
the structure or the symmetry  of the quantum spectrum.
In this case we expect the
Bargmann angle to be time periodic with  typical
 quantum revivals. In contrast, in the
 classical  chaotic regime, $\hat{V}(t)$
changes qualitative the quantum  spectrum.

 Starting from
the (classically) regular case we conclude  that, if pair excitations  are
neglected, then the energy spectrum of the unperturbed  part has the form

\begin{equation}
E_{n}=(n-1-N/2)H_{0z}+\frac{1}{4}J(N-4n+3)+\frac{\beta N}{4}.
\end{equation}
With this spectrum we infer for the Bargmann angle (eq.(57)) the expression
\begin{equation}
A^{2}(t)=\frac{1}{N^{2}}\bigg( \frac{\sin(N+1/2)t-\sin(t/2)}{2\sin(t/2)}\bigg)^{2}+\frac{1}{N^{2}}\bigg( \frac{\cos(t/2)-\cos(N+1/2)t}{2\sin(t/2)}\bigg)^{2}.
\end{equation}
As evident  from Fig.11, if the underlying classical dynamics is   regular,
the    time dependence of the Bargmann angle is periodical and  the system is
characterized by quantum revivals.

To deal quantum mechanically with the classically  chaotic regime
we follow Ref. \cite{Haake} and employ a
 Gaussian orthogonal ensemble \cite{Haake}.
\begin{equation}
P(\varphi_{1},\ldots\varphi_{N})=\prod\limits_{n>m}
(\varphi_{n}-\varphi_{m})\exp\bigg(-\sum\limits_{n=1}^{N}\varphi_{n}^{2}\bigg).
\label{eq:gaus}
\end{equation}
We note here that in general the distribution function (60)
includes correlations between all $N$ levels.
If the  number of
correlated levels is $n$, then  the  $n<N$ level-correlated
distribution function reads
\begin{equation}
P_{n}(\varphi_{1},\ldots\varphi_{n})=\frac{N!}{(N-n)!}\int
P_{n}(\varphi_{1},\ldots \varphi_{N})d\varphi_{n+1}\ldots
d\varphi_{N}. \label{eq:gaus1}
\end{equation}
The structure of the expression (57) suggests that
the second-order correlated level distribution function
$P_{2}(\varphi_{n},\varphi_{m})$ is sufficient (each term in the
sum contains two phases).
 Upon straightforward calculations we reduced
$P_{2}(\varphi_{n},\varphi_{m})$  to
$$
P_{2}(\varphi_{n},\varphi_{m})=K_{N}(\varphi_{n},\varphi_{n})K_{N}(\varphi_{m},\varphi_{m})-K_{N}(\varphi_{n},\varphi_{m})K_{N}(\varphi_{m},\varphi_{n})
,$$ with
 $$
K(\varphi_{n},\varphi_{m})=\sum\limits_{k=1}^{N}\phi_{k}(\varphi_{n})\phi_{k}(\varphi_{n}),
$$ and $$
\phi_{k}(\varphi)=\frac{1}{\big(2^{n}n!\sqrt{\pi}\big)^{1/2}}H_{k}(\varphi)
\exp\bigg[-\frac{\varphi^{2}}{2}\bigg]
.$$
  $H_{n}(\varphi)$ are Hermite polynomials. For $\big\langle A^{2}(t)\big\rangle_{P_{2}}$  we find 
(for $N\gg1$)
\begin{eqnarray}
\big\langle
A^{2}(t)\big\rangle_{P_{2}}=C\exp\big[-t^{2}/2\big]
\bigg\{\sum\limits_{n,m}^{N}L_{n}^{0}\big(t^{2}/2\big)L_{m}^{0}
\big(t^{2}/2\big)-\frac{\sqrt{\pi}}{4}\sum\limits_{m=1}^{N}
\sum\limits_{n=1}^{N}\frac{n!}{2^{m-n}m!}t^{m-n}
\nonumber \\
\big(L_{n}^{m-n}\big(t^{2}/2\big)\big)^{2}-\frac{\sqrt{\pi}}{4}
\sum\limits_{m=1}^{N}\sum\limits_{n=1}^{m-1}\frac{n!}{2^{m-n}m!}t^{m-n-1}
\big(L_{n}^{m-n-1}\big(t^{2}/2\big)\big)^{2}\bigg\}
\label{eq:barg}\end{eqnarray} where $L_{n}^{m}(t^{2}/2)$ are
Laguerre polynomials. The constant  $C$ we find from the condition  $\big\langle
A^{2}(0)\big\rangle_{P_{2}}=1$.
 Dynamical freezing (DF) means then a stabilization over time of the quantum
  distance \cite{Matos-Abiague2,Matos-Abiague} quantified by the
  Bargmann angle. To test for this situation we  numerically solve for (62);
  a typical example is shown in Fig.(\ref{Fig:14}).
  These calculations are performed for parameters appropriate for the classically chaotic regime,
  e.g. those of Fig.\ref{Fig:2}.  The interpretation of
  Fig.(\ref{Fig:14}) is that  SC drives the system diffusively to
 the  target state $\big|\Psi(t)\big\rangle$, which in this case is
 orthogonal to
$\big|\Psi(0)\big\rangle$. The Bargmann angle is therefore deflected within
a time $t_D$ determined by the diffusion constant  to the
value $\theta_{B}(t_D)\sim\pi/2$. DF is then evidenced by a small variations
of $\theta_{B}(t_D)$ for $t>t_D$.
\subsection{Conclusions}
For an  exchange-coupled, non-linear spin chain
and in the presence of a (uniaxial) magnetic anisotropy and external driving fields,
 stochastic switching is possible if the field parameters are chosen such that the
 underlying classical dynamics is chaotic. The switching mechanism is identified to be the
 Arnold-type diffusion.  This we concluded analytically and substantiated
 by full   numerical semi-classical and quantum calculations.
We also inspected  the possibility of dynamical freezing, i.e. stabilizing
 the target state beyond the switching time.

\textbf{Acknowledgment:} The  project is financially supported by
 the Georgian National Foundation
(grants: GNSF/STO 7/4-197, GNSF/STO 7/4-179). The financial
support by the Deutsche Forschungsgemeinschaft (DFG) through SFB
762 and though  SPP 1285 is gratefully acknowledged.
\appendix
\section{}
The canonical Lyapunov system  has the general  structure
\begin{eqnarray} \label{eq:A1}
&& \dot{x}=-\lambda y+X(x,y,z_{1}\ldots z_{m});\nonumber \\
&& \dot{y}=\lambda x+Y(x,y,z_{1}\ldots z_{m}); \\
&& \dot{z}_{s}=\sum\limits_{j=1}^{m}b_{sj}+Z_{s}(x,y,z_{1}\ldots
z_{m}); \nonumber \\
 && (s=1,2,\ldots m; m=n-2)\nonumber
\end{eqnarray}
We seek a reduction  of this  system of equations  using the canonical
transformation
\begin{eqnarray}
&& \dot{x}=-\delta y-\gamma y z, \nonumber \\
&& \dot{y}=\delta x-z+\gamma x z,\\
&& \dot{z}=y. \nonumber \label{eq:A2}
\end{eqnarray}
We are interested in the linear part of   eq. (\ref{eq:A2}), i.e.
\begin{eqnarray} \label{eq:A3}
&& \dot{x}=-\delta y,\nonumber \\
&& \dot{y}=\delta x-z,\\
&& \dot{z}=y,\nonumber
\end{eqnarray}
 From the coefficients of the equation (\ref{eq:A3}) we can construct the following matrix
\begin{equation} \label{eq:A4}
 a= \left( \begin{array}{c}
  0~~-\delta~~0 \\
\delta~~0~~-1 \\
  0~~1~~0 \end{array} \right)
.\end{equation}
 We consider now the linear transformation
\begin{eqnarray}\label{eq:A5}
&&\xi_{1}=\gamma_{11}x+\gamma_{12}y+\gamma_{13}z,\nonumber \\
&&\xi_{2}=\gamma_{21}x+\gamma_{22}y+\gamma_{23}z,\\
&&\xi_{3}=\gamma_{31}x+\gamma_{32}y+\gamma_{33}z.\nonumber
\end{eqnarray}
In the new variables  eq. (A3) is cast as
\begin{equation}\label{eq:A6}
\frac{d\xi_{i}}{dt}=\lambda_{i}\xi_{i}~~~~i=1,2,3
.\end{equation}
Taking eqs.(\ref{eq:A5}), (\ref{eq:A3}) into account we infer  from eq.
(\ref{eq:A6}) that
\begin{eqnarray}\label{eq:A7}
&&(a_{11}-\lambda_{i})\gamma_{i1}+a_{21}\gamma_{i2}+a_{31}\gamma_{i2}=0,\nonumber
\\
&&a_{12}\gamma_{i1}+(a_{22}-\lambda_{i})\gamma_{i2}+a_{32}\gamma_{i3}=0,\\
&&a_{13}\gamma_{i1}+a_{23}\gamma_{i2}+(a_{33}-\lambda_{i})\gamma_{i3}=0.
\nonumber
\end{eqnarray}
Equating the  determinant to zero
\begin{equation} \label{eq:A8}
  \left| \begin{array}{c}
 -\lambda_{i}~~-\delta~~0 \\
  \delta ~~-\lambda_{i}~~-1 \\
  0  ~~1~~ -\lambda_{i} \end{array} \right|=0,
\end{equation}
 we find
\begin{equation} \label{eq:A9}
 \lambda_{1,2}=\pm i \lambda,~~ \lambda_{3}=0,~~
\lambda=\sqrt{1+\delta^{2}}.
\end{equation}
 According to (\ref{eq:A7}) this gives the following solutions for the
 matrix
\begin{eqnarray} \label{eq:A10}
&&\gamma_{11}=-\delta
,~~-\gamma_{12}=-i\lambda,~~\gamma_{13}=1;\nonumber \\
&&\gamma_{21}=-\delta, ~~-\gamma_{22}=-i\lambda,~~\gamma_{23}=1;\\
&&\gamma_{31}=-
\frac{1}{\delta},~~-\gamma_{22}=0,~~\gamma_{33}=1.\nonumber
\end{eqnarray}
 So the canonical transformation (\ref{eq:A5}) has the  form
\begin{eqnarray} \label{eq:A11}
&& \xi_{1}=-\delta x-i\lambda y+z; \nonumber \\
&& \xi_{2}=-\delta x+i\lambda y+z;\\
&& \xi_{3}=-\frac{1}{\delta} x+z.\nonumber
\end{eqnarray}
 The inverse transformation reads
\begin{eqnarray}\label{eq:A12}
&&
x=\frac{\delta}{\lambda^{2}}\bigg(\xi_{3}-\frac{\xi_{1}+\xi_{2}}{2}\bigg),\nonumber
\\
&& y=\frac{i}{2\lambda}(\xi_{1}+\xi_{2}),\\
&&z=\frac{\xi_{1}+\xi_{2}+2\delta^{2}\xi_{3}}{2\lambda^{2}}.\nonumber
\end{eqnarray}
Using eq.(\ref{eq:A10}) we obtain then for equations of motion in the variable $\xi_{i}$
\begin{eqnarray}\label{eq:A13}
&& \dot{\xi}_{1}=i\lambda\xi_{1},\nonumber \\
&& \dot{\xi}_{2}=-i\lambda\xi_{2},\\
&& \dot{\xi}_{3}=0. \nonumber
\end{eqnarray}
To reduce our system of equations to the canonical form, one more
transformation is needed
\begin{eqnarray} \label{eq:A14}
 && x_{1}=-(\xi_{1}+\xi_{2}), \nonumber \\
&& y_{1}=i(\xi_{1}-\xi_{2}), \\
&& z_{1}=\xi_{3}.\nonumber
\end{eqnarray}
Taking  eqs.(\ref{eq:A10}) , (\ref{eq:A11}) into account  we
obtain
\begin{eqnarray}\label{eq:A15}
&& x_{1}=2\delta x-2z; \nonumber \\
&& y_{1}=2\lambda y;\\
&& z_{1}=\frac{1}{\delta} x+z. \nonumber
\end{eqnarray}
The inverse transformation reads
\begin{eqnarray}\label{eq:A16}
&&
x=\frac{\delta}{\lambda^{2}}\bigg(z_{1}+\frac{x_{1}}{2}\bigg),\nonumber
\\
&& y=\frac{1}{2\lambda}y_{1},\\
&&
z=\frac{1}{\lambda^{2}}\big(-x_{1}+2\delta^{2}z_{1}\big).\nonumber
\end{eqnarray}
 Using
(\ref{eq:A15}), (\ref{eq:A16}) we can finally  rewrite the set of equations in
the canonical form
\begin{eqnarray}\label{eq:A17}
&&\dot{x_{1}}=-\lambda
y_{1}+\frac{\delta\gamma}{2\lambda^{3}}y_{1}x_{1}-\frac{\gamma\delta^{3}}{\lambda^{3}}y_{1}z_{1},\nonumber
\\
&&\dot{y_{1}}=\lambda
x_{1}+\frac{\gamma\delta(\delta^{2}-1)}{\lambda^{3}}x_{1}z_{1}-\frac{\gamma\delta}{2\lambda^{3}}x_{1}^{2}+\frac{2\gamma\delta^{3}}{\lambda^{3}}z_{1}^{2},\\
&&\dot{z_{1}}=\frac{\gamma}{4\lambda^{3}\delta}y_{1}x_{1}-\frac{\gamma\delta}{2\lambda^{3}}y_{1}z_{1}.
\nonumber
\end{eqnarray}


\begin{thebibliography}{29}
 %
 \bibitem{science} C. F. Hirjibehedin, C. P. Lutz, and A. J. Heinrich, Science \textbf{312}, 1021 (2006);
 S. Rusponi \emph{et al.}, Nature Mater. \textbf{2}, 546 (2003);
 T. Mirkovic \emph{et al.}, Nature Nanotech. \textbf{2}, 565 (2007);
 J. A. Stroscio and R. J. Celotta, Science \textbf{306}, 242 (2004).
 %
 \bibitem{spintheory} S. Lounis, Ph. Mavropoulos, P. H. Dederichs, and S. Bl\"ugel, Phys. Rev. B \textbf{72}, 224437 (2005); S. Lounis, P. H. Dederichs, and S. Bl\"ugel, Phys. Rev. Lett. \textbf{101}, 107204 (2008);
     P. Politi and M. Gloria Pini
\textbf{79}, 012405 (2009) and references therein.
 %
\bibitem{Abragam} A.  Abragam and M. Goldman Nuclear
Magnetism: Order and Disorder (Oxford: Oxford
     University Press) (1982)
 %
\bibitem{Mejia-Monasterio}  C. Mejia-Monasterio, T. Prosen, and G. Gasati,
Europhys.Lett. 72 (4), p.520, (2005)
%
\bibitem{Saito}  K. Saito,Europhys.Lett. 61, 34 (2003)
%
\bibitem{Sagdeev}  G.M. Zaslavsky, \emph{The physics of chaos
in hamiltonian systems} 2ed edition, (Imperial College London, 2007).

 %
\bibitem{Ckhvaradze} L. L.Chotorlishvili, V. M. Ckhvaradze  Low.Temp.Physics v. 30, No. 9,
p. 739 (2004)
%
\bibitem{Matos-Abiague} A. Matos-Abiague and J. Berakdar Phys.Rev A 73, 024102 (2006)
 %
\bibitem{Matos-Abiague2} A. Matos-Abiague and J. Berakdar Euro. Phys. Lett 71,  705-711 (2005)
 %
 \bibitem{Yuan} Xiao-Zhong Yuan, Hsi-Sheng Goan, and Ka-Di Zhu, Phys.Rev. B 75, 045331 (2007).
 %
 \bibitem{Tuchette} Q.A. Tuchette et al., Phys.Rev.Lett. 75, 4710 (1995).
 %
 \bibitem{Mabuchi} H. Mabuchi and A. Doherty, Science 298, 1372 (2002).
 %
\bibitem{Hood}  C.J. Hood et al., Science 287, 1447 (2000).
%
 \bibitem{Raimond}  J.Raimond,M.Brune, and S.Haroche, Rev.Mod.Phys. 73, 565 (2001).
%
\bibitem{Chotorlishvili} L.Chotorlishvili, Z. Toklikishvili, Phys.Lett.A v. 372, 2806 (2008)
%
\bibitem{Lakshmanan}  M. Lakshmanan and A. Saxena arXiv:
0712.2503v1, (2007)
%
 \bibitem{Zolotaryuk}  Y. Zolotaryuk, S. Flach, and V. Fleurov,
Phys.Rev.B, 214422, (2001)
%
\bibitem{prl09} A. Sukhov,  J. Berakdar Phys.Rev.Lett. \textbf{102}, 057204 (2009)
%
\bibitem{Balescu} R. Balescu. Equilibrium and Nonequilibrium Statistical
Mechanics. (New York, Wiley, 1975.)
%
\bibitem{Lichtenberg} A. J. Lichtenberg and M.A. Lieberman, Regular and Chaotic Dynamics (Springer-Verlag, New
York, 1992),
%
\bibitem{Haken} H. Haken, Synergetics. An Introduction, Berlin:
Springer-Verlag, 1978.
%
\bibitem{levanpla} L.Chotorlishvili, Z. Toklikishvili, J. Berakdar Phys.Lett.A  \textbf{373}, 231 (2009)
%
\bibitem{Chirikov}  B. V. Chirikov. Physics Reports, Volume 52, Issue 5, May 1979,
Pages 263-379
%
\bibitem{Kittel} Ch. Kittel, Introduction to Solid State Physics, Fourth Edition, John Wiley and
Sons, (New York, London, Sydney, Toronto, 1978),
%
\bibitem{Ugulava} A. I. Ugulava, L. L. Chotorlishvili, Z. Z. Toklikishvili, and A. V.
Sagaradze, Low.Temp.Physics, v.32, No. 10, p. 915, (2006)
%
\bibitem{Pati} A. K. Pati and S. V. Lawande, Phy. Rev. A v.58, No. 2, 831,
(1998)
%
\bibitem{Anandan} J. Anandan, Y. Aharonov Phys.Rev.Lett. v.65, N. 14, p. 1697
(1990)
%
\bibitem{vonBalk} R. von Balk, Eur. J Phys 11 215-220, (1990)
%
\bibitem{Stockman} H.J. St\"ockmann, Quantum Chaos, An Introduction (Cambridge
Univ.Press, Cambridge, 1993).
%
\bibitem{Haake} F.Haake, Quantum Signatures of Chaos (Springer, Berlin,2001).
%
\bibitem{Skrinnikov} L. Chotorlishvili, V. Skrinnikov Phys. Lett. A 372, 761-768,
(2008)
%
%
\end{thebibliography}
\end{document}